\documentclass{aa}  

\usepackage{graphicx}
\usepackage{txfonts}
\usepackage{natbib}
\usepackage{threeparttable}
\usepackage{caption}
\usepackage{subcaption}
\usepackage{color}
\usepackage[colorlinks=true,linkcolor=blue]{hyperref}

\usepackage{url}
\usepackage{multirow}
\usepackage{comment}

\usepackage[normalem]{ulem}

\begin{document}

\title{Dwarf galaxies as a probe of a primordially magnetized Universe}

   \author{Mahsa Sanati\inst{\ref{epfl}},\inst{\ref{oxford}}, Sergio Martin-Alvarez\inst{\ref{kipac}}, Jennifer Schober\inst{\ref{epfl}}, Yves Revaz\inst{\ref{epfl}}, Adrianne Slyz\inst{\ref{oxford}}, Julien Devriendt\inst{\ref{oxford}}}

    \institute{Subdepartment of Astrophysics, University of Oxford, Keble Road, Oxford OX1 3RH, UK \label{oxford} \\
        \email{mahsa.sanati@physics.ox.ac.uk}
    \and  Kavli Institute for Particle Astrophysics \& Cosmology (KIPAC), Stanford University, Stanford, CA 94305, USA
    \label{kipac}
    \and Institute of Physics, Laboratory of Astrophysics, \'{E}cole Polytechnique F\'{e}d\'{e}rale de Lausanne (EPFL), 1290 Sauverny, Switzerland \label{epfl}
        }

   \date{Received: XXXX; accepted: YYYY}

  \abstract
    { 
    {\textit{Aims}}: The true nature of primordial magnetic fields (PMFs) and their role in the formation of galaxies still remains elusive. To shed light on these unknowns, we investigate their impact by varying two sets of properties: (i) accounting for the effect of PMFs on the initial matter power spectrum, and (ii) accounting for their magneto-hydrodynamical effects on the formation of galaxies. By comparing both we can determine the dominant agent in shaping galaxy evolution.
    
    \textit{Methods:} We use the magneto-hydrodynamics code \texttt{RAMSES}, to generate multiple new zoom-in simulations for eight different host halos of dwarf galaxies across a wide luminosity range of $10^3-10^6\,L_{\odot}$. These halos are selected from a $\Lambda$CDM cosmological box, tracking their evolution down to redshift $z=0$. We explore a variety of primordial magnetic field (comoving) strengths $B_{\lambda}$ ranging from $0.05$ to $0.50\,\mathrm{nG}$.

    \textit{Results:}
    We find magnetic fields in the interstellar medium not only modify star formation in dwarf spheroidal galaxies but also completely prevent the formation of stars in less compact ultra-faints with halo mass and stellar mass below $\sim 2.5\cdot10^9$ and $3\cdot10^6\,M_{\odot}$, respectively. At high redshifts, the impact of PMFs on host halos of dwarf galaxies through the modification of the matter power spectrum is more dominant than the influence of magneto-hydrodynamics in shaping their gaseous structure. Through the amplification of small perturbations ranging in mass from $10^7$ to $10^9\,M_{\odot}$ in the $\Lambda$CDM$+$PMFs matter power spectrum, primordial fields expedite the formation of the first dark matter halos, leading to an earlier onset and a higher star formation rate at redshifts $z>12$. 
    We investigate the evolution of various energy components and demonstrate that magnetic fields with an initial strength of $B_{\lambda}\geq0.05\,\mathrm{nG}$ exhibit a strong growth of magnetic energy, accompanied by a saturation phase, that starts quickly after the growth phase. These trends persist consistently, regardless of the initial conditions, whether it is the classical $\Lambda$CDM or modified by PMFs. Lastly, we investigate the impact of PMFs on the present-time observable properties of dwarf galaxies, namely, the half light radius, V-band luminosity, mean metallicity and velocity dispersion profile. We find that PMFs with moderate strengths of $B_{\lambda}\leq0.10\,\mathrm{nG}$ show great agreement with the scaling relations of the observed Local group dwarfs. However, stronger fields lead to large sizes and high velocity dispersion. 
   }
  
   \keywords{Primordial magnetic fields, Magnetic Fields, Cosmology, Dwarf galaxy, Ultra faint dwarf galaxies, Galaxy evolution, Star Formation, Magnetohydrodynamical simulations}

   \titlerunning{XXX}   

   \maketitle

%

\section{Introduction}\label{sec:intro}



Observations reveal the pervasive nature of magnetic fields in our Universe, present at all cosmic scales probed so far; be it the small scale of planets, stars and galaxies \citep{2010SSRv..152..651S,Reiners2012,2001SSRv...99..243B, 2013pss5.book..641B}, 
or at the large scale of galaxy clusters \citep{2001ApJ...547L.111C, 2004IJMPD..13.1549G, 2005A&A...434...67V}. 
In particular for galaxies, magnetic fields play an important role in the process of star formation 
\citep{2011ApJ...730...40P,2018MNRAS.474.4824Z,KrumholzFederrath2019}. 
They have the potential to increase the gas fragmentation \citep{2019MNRAS.485.3024I}, slow down the circumgalactic gas flows \citep{2021MNRAS.501.4888V}, suppress the recycling outflows \citep{2018ApJ...865...64G}, and shrink the size of galaxies \citep{2020MNRAS.495.4475M}. 
The role of magnetic fields in shaping the complex multi-phase interstellar medium is widely recognized \citep{2017A&A...604A..70I, 2019MNRAS.489.5004K}. Observational data from neighboring galaxies indicate that magnetic energy permeates all phases of the interstellar medium and is approximately in equipartition with the thermal and kinetic energy components \citep{2015A&A...578A..93B, 2023ApJ...942L..13L}. 

Various theories have been proposed to explain the existence of strong galactic fields with $\mu\mathrm{G}$ amplitude 
that are coherent on $\mathrm{kpc}$ scales and are detected through radio spectropolarimetry in spiral and ultraluminous infrared galaxies \citep{2011MNRAS.412.2396F,2015A&ARv..24....4B, 2008ApJ...680..981R, 2013ApJ...763....8M}, and are even present in high-redshift galaxies 
\citep{2008Natur.454..302B, MaoEtAl2017, GeachEtAl2023}.
Amongst them, dynamo mechanisms 
in collapsed objects 
\citep{2006AstHe..99..568I, 2008Sci...320..909R, 2013PhRvL.111e1303N, SchoberEtAl2013, 2022MNRAS.513.3326M}, as well as 
magnetized flows in  
supernovae explosions \citep{2002RvMP...74..775W, 2005ApJ...633..941H, 2018MNRAS.479..315S} are only a channel of amplification and 
require a seed magnetic field.
During the process of galaxy formation, the collapse of gas can result in the entanglement of magnetic flux, which may also serve as another means of amplification and lead to
galactic fields \citep{2012MNRAS.423.3148S}.
However, in this case, a strong micro gauss initial seed, comparable to what is observed today, is required.

Seed magnetic fields for the galactic dynamo can be generated by specific 
astrophysical mechanisms, like the Biermann battery
during structure formation \citep[e.g.][]{biermann_kulsrud_1997, biermann_ryu_1998} or reionization \citep[e.g][]{biermann_subramanian_1994,
gnedin_2000_reionization} or the Weibel instability \citep{SironiEtAl2023,ZhouEtAl2024}.
Alternatively,
cosmic magnetic fields can be remnants from the early Universe, generated 
in the course of phase transitions 
\citep{1983PhLB..133..172H, 1992ApJ...391L...1R,ellis_2019_phase_trans_bfields}
or during the inflationary expansion \citep{1988PhRvD..37.2743T, 2016JCAP...10..039A, 2019JCAP...10..032D, 2019JCAP...09..008F}. Inflation, in particular, provides an ideal mechanism for producing large-scale magnetic fields \citep{ratra1992, turner1988}.
These fields could undergo further 
amplification through a small-scale chiral dynamo
\citep{JoyceShaposhnikov1997, 2022PhRvD.105d3507S}. 
Such primordial magnetic fields, when generated during this early epoch, would be amplified during the collapse of density perturbations, and directly explain the magnetic fields observed in galaxies \citep{2011PhR...505....1K, 2021MNRAS.504.2517M}.

Observational evidence of non-negligible magnetism even in the intergalactic medium voids has recently sparked interest in this intriguing possibility 
\citep{2010Sci...328...73N, 2011A&A...529A.144T, 2011MNRAS.414.3566T}.
These observations are based on
a missing $\gamma$-ray signal from TeV blazars located behind the voids. 
The presence of magnetic fields in these relatively quiet environments is difficult to explain purely by an astrophysical process in the late Universe \citep{2006MNRAS.370..319B,2001ApJ...556..619F, 2017CQGra..34w4001V,  2020MNRAS.495.2607O, 2021MNRAS.505.5038A, TjemslandEtAl2023}, and would perhaps favor a primordial origin. 
The recent discovery of magnetized filaments between 
galaxy clusters reinforces this scenario 
\citep{2019Sci...364..981G, 2022MNRAS.512..945C}. 
For comprehensive reviews on different scenarios for the generation of magnetic fields see \citet{2002RvMP...74..775W}, \citet{2011PhR...505....1K}, and \citet{2016RPPh...79g6901S}.

The main interest in magneto-genesis scenarios with a primordial origin 
stems from their capacity in simultaneously explaining the $\gtrsim10^{-16}\,\mathrm{G}$ intergalactic magnetic fields as well as their
$\sim10^{-5}\,\mathrm{G}$ 
galactic counter-parts without imposing a time 
constraint 
for rapid amplification and subsequent reorganization into their observed small and large-scale ISM distribution (e.g., \citealt{2015A&A...578A..93B}; \citealt{2023ApJ...952....4B}).
The existence of such magnetic fields in the 
pre-recombination plasma may even help to alleviate the Hubble tension \citep{2020PhRvL.125r1302J}. 
This result emerges from an increased rate of recombination in the presence of strong primordial magnetic fields \citep{2022PhRvD.105b3513G}. Before and during the Epoch of Recombination, these primordial magnetic fields generate density perturbations in addition to those produced by inflation.
%
These perturbations are mapped on the last scattering surface of the cosmic microwave background (CMB) photons, where primordial magnetic fields can imprint a variety of signals.
The trace of a stochastic primordial magnetic field on the CMB map could be observed in the excess temperature and polarization anisotropies \citep{2006MNRAS.366.1437S, 2007NewAR..51..275D, 2012SSRv..166...37W, 2016A&A...594A..13P}.
These magnetically-produced anisotropies display a strongly non-Gaussian pattern 
\citep{2009JCAP...12..024C, 2009PhRvL.103h1303S,2014PhRvD..89d3523T}.
A comprehensive analysis of the CMB constrains the primordial magnetic field to have an upper limit for its amplitude of a few nano Gauss, favouring a nearly scale invariant spectrum smoothed over the spatial scale of $1\,\mathrm{Mpc}$.
Alternative observational constraints of primordial magnetic fields stem from 
Big Bang nucleosynthesis \citep[][which provides the strongest constraints on all length scales]{KernanEtAl1996},
the Sunyaev-Zel'dovich statistics \citep{tashiro2009}, the propagation of ultra-high energy cosmic rays 
\citep{2018ApJ...861....3B,AlvesBatistaSaveliev2019},
the two-point shear correlation function from gravitational lensing \citep{2012ApJ...748...27P}, and Lyman-$\alpha$ forest clouds \citep{2013ApJ...762...15P}.
Density perturbations produced by primordial magnetic fields influence the formation of first structures and the total matter power spectrum.  
This impact commences when the ionized matter is decoupling from photons \citep{2012PhRvL.108w1301T}. At this stage, magnetic Alfv\'en waves can produce additional motions in the nearly homogeneous background
\citep{2000PhRvD..61h3519T}. These additional density and velocity perturbations in the baryonic matter \citep{1996ApJ...468...28K, 1978PhDT........95W} are gravitationaly coupled to dark matter over-densities, and consequently modify 
the number distribution of the first dark matter halos \citep{2017Ap&SS.362...16V, 2018Ap&SS.363...93C}.
The magnetically-produced perturbations dominate the standard $\Lambda$ cold dark matter ($\Lambda$CDM) power spectrum at large $k$-modes \citep{1978PhDT........95W, 2003JApA...24...51G, 2012PhRvD..86d3510S}, and therefore affect the formation of small scale structures. 
This region of $k$-space encompasses the low mass dark matter halos, hosting the smallest galactic systems, dwarf galaxies %
\citep[see][for a recent review]{2019ARA&A..57..375S}.

Despite being the most abundant type of galaxies, dwarf galaxies are faint and difficult to observe. Therefore, the best studied population of dwarf galaxies have been discovered in our proximity \citep[][updated online catalog]{2012AJ....144....4M}, and in the nearby Virgo cluster \citep{1985AJ.....90.1681B}.
With the new window the James Webb Space Telescope (JWST) opens to the young Universe, high redshift dwarf galaxies further expand our observational data \citep{2019MNRAS.485.5939J, 2021ApJ...913L..25G, 2022arXiv221112970Y}.
The existing observations reveal the scaling of dwarf properties, from brighter systems, to the regime of ultra faint dwarfs. 
Some scaling relations that characterize the properties of dwarf galaxies are the size-luminosity \citep[e.g.][]{2018ApJ...866L..21P}, metallicity-luminosity \citep{lequeux1979, Skillman1989, Garnett2002, Tremonti2004}, and luminosity-velocity dispersion \citep[e.g.][]{battaglia2008, 2009AJ....137.3100W, 2011EAS....48...81F, fabrizio2016}. 
Thus far cosmological simulations have been relatively successful in reproducing the scaling relations for dwarf spheroidals
\citep{jeon2017, Maccio2017, Revaz2018, Escala2018,Wheeler_2019, agertz2020, applebaum2020, prgomet2022}.
However, the assembly of fainter galaxies within dark matter halos is still subject to persistent tensions in a $\Lambda$CDM paradigm.
The most prominent are the diversity of sizes \citep{2015MNRAS.452.3650O, 2019ApJ...887...94R,2023A&A...679A...2R} and diversity of rotation curves \citep{2015MNRAS.452.3650O}, the satellite planes \citep{2013MNRAS.435.2116P}, and the too-big-to-fail problem \citep{boylankolchin2011,boylankolchin2012}. For a recent review on the long-standing tensions between $\Lambda$CDM and Local Group observations see \citet{2017ARA&A..55..343B} and \citet{2022NatAs...6..897S}.     

Modern cosmological simulations have made significant progress in addressing small-scale challenges in the standard model by incorporating more precise treatments of baryonic physics \citep[e.g.,][]{2021MNRAS.501.5597G, 2022MNRAS.516.2112K, 2023MNRAS.525.3806M, 2023MNRAS.519.3154H}. The importance of magnetic fields in shaping galaxies and their the interstellar medium has been recognized and studied by a variety of simulations \citep[][]{2020MNRAS.495.4475M,2017MNRAS.469.3185P, 2020MNRAS.491.1190S}. However, accounting for additional factors such as primordial magnetic fields, which are often ignored in models, will provide new insights into existing tensions.
Presenting a suite of cosmological simulations, \citet{sanati2020} showed the impact of primordial magnetic fields on the formation of dwarf galaxies during the Epoch of Reionization. 
The aim of this manuscript is to expand on that work, by investigating the role played by primordial magnetic fields in the formation of dwarf galaxies and shaping their global properties, conducting a set of cosmological zoom-in simulations 
that are designed to self-consistently model: (i) the formation and evolution of dwarf galaxies emerging from a magnetically-distorted $\Lambda$CDM matter power spectrum,
(ii) the growth and amplification of uniform primordial magnetic fields.

This paper is organized as follows.
Section~\ref{sec:primordial_magnetic_field} reviews 
the role played by primordial magnetic fields in the early history of the Universe, and their influence on the $\Lambda$CDM matter power spectrum. 
The numerical framework to generate and evolve our simulations is described in details in Section~\ref{sec:simulations}.
The results are presented in Section~\ref{sec:results}.
We first discuss the first stages in the evolution of dwarf galaxies, including the first star forming gas halos in Section~\ref{sec:dens}. Then in Section~\ref{sec:energy} we present the evolution of different energy components. In Section~\ref{sec:sfr} we follow the star formation history and the scaling relations of our simulated dwarf models.  
Finally, a summary of our main conclusions and a brief discussion is presented in Section~\ref{sec:conclusions}.

%


\section{Inclusion of primordial magnetic fields in $\Lambda$CDM}\label{sec:primordial_magnetic_field}


 In Section~\ref{sec:intro}, it was discussed that cosmic magnetic fields may have originated from the inflationary epoch 
 or phase transitions in the early Universe. 
 Such primordial 
 fields with a stochastic nature modify the baryonic clumping factor, thereby affect the formation of initial structures and the $\Lambda$CDM matter power spectrum. This section provides a detailed review of such effects.

\subsection{Primordial magnetic field power spectrum}

Primordial magneto-genesis scenarios motivated by inflation, conventionally result in a homogeneous, isotropic and Gaussian random initial magnetic field $\mathrm{\vec{B}}\,(x,t)$ \citep{wasserman1978, 1996ApJ...468...28K, 2003JApA...24...51G}. The two-point correlation function of 
a non-helical 
magnetic field $\mathrm{\vec{B}}$ in Fourier space, therefore, can be described by the relation \citep{1987flme.book.....L}:   
\begin{equation}
    \langle B_i(\vec{k})\, B_j^*(\vec{k}') \rangle = (2\pi)^3\, \delta(\vec{k}-\vec{k}') \frac{P_{ij}(k)}{2} P_{B}(k)  ,
\end{equation}
where $k = |\vec{k}|$ is the comoving wave number. 
The power spectrum of the primordial magnetic field 
$P_{B}(k)$ is defined by a simple power law: 
\begin{equation}
    P_B(k) = A_{B}\,k^{n_B},
    \label{eq:pb}
\end{equation}
and projected into the Fourier space, orthogonal to $k_i$, through the transverse projection tensor\footnote{The importance of the projection tensor lies in the fact that estimating the perpendicularly projection to the line of sight is easier than measuring the tensor field from the observations in large-scale structure.} $P_{ij}(k)=\delta_{ij}-k_i k_j/k^2$. 
Here $n_B$ is the slope, and $A_B$ the amplitude of the magnetic power spectrum. 
Commonly, the amplitude of the spectrum is defined as the variance of magnetic field strength at the present time $\mathrm{\vec{B}}_{\lambda}\,(x,t_0)$, smoothed at scales of $\lambda=2\pi/k_{\lambda}=1\,\mathrm{Mpc}$, as
\begin{eqnarray}
    \langle B^2_{\lambda}\,(x,t_0) \rangle =
    \frac{1}{\pi^2}\int_{0}^{k_c}\mathrm{d}\vec{k}\, \vec{k}^2\, (2\pi)^3\, \delta(\vec{k}-\vec{k}') \frac{P_{ij}(k)}{2} P_{B}(k), 
    \label{eq:B2}
\end{eqnarray}
where the integral upper limit $k_c = 1\,\mathrm{Mpc}^{-1}$. From Eq.~\ref{eq:B2} the amplitude $A_B$ is expressed as  \citep{2012PhRvD..86d3510S}
\begin{equation}
    A_B = \frac{(2\pi)^{n_B+5}\,B^2_{\lambda}}{2\Gamma\,(\frac{n_B+3}{2})\,k^{n_B+3}_\lambda}.
    \label{eq:pb}
\end{equation}
With this definition, $n_B$ and $B_\lambda$ fully characterize the primordial magnetic fields. They constitute our main free parameters. 
In the following section, we will describe the incorporation of primordial magnetic fields into the evolution of density perturbations using the linearized Newtonian theory.

\subsection{Evolution of density fluctuations}\label{perturbation_growth}

In the $\Lambda$CDM framework, all cosmic structures are the result of primordial quantum field fluctuations.   
In the early Universe, the causal connection between these small over-densities shapes a nearly scale invariant power spectrum $P(k)\propto k^n$, as a function of wave number $k$, with $n\simeq1$ \citep{1970PhRvD...1.2726H, 1972MNRAS.160P...1Z, 1970ApJ...162..815P, 2016A&A...594A..13P}. 

In the linear Newtonian paradigm, the evolution of initial density perturbations in the baryonic matter $\delta_b(\vec{x},t) = \delta \rho_b(\vec{x},t)/{\rho}_b$ 
and the collisionless dark matter $\delta_{\rm{DM}}(\vec{x},t)= \delta \rho_{\rm{DM}}(\vec{x},t)/{\rho}_{\rm{DM}}$ is described by the two following coupled equations \citep[see, e.g.,][]{dodelson2003modern},
%
\begin{eqnarray}
      \frac{\partial^2 \delta_b}{\partial t^2}   + \left[2\,H(t)  + \frac{4\rho_{\gamma}}{3 \rho_b} n_e \sigma_T a   \right] \frac{\partial \delta_b}{\partial t} & = & \nonumber \\
      c_b^2 \nabla^2 \delta_b +  4\pi G \left[ {\rho}_b \delta_b + {\rho}_{\rm{DM}} \delta_{\rm{DM}} \right]
  \label{eq:deltab}
\end{eqnarray}
and
\begin{equation}
      \frac{\partial^2\delta_{\mathrm{DM}}}{\partial t^2}  + 2\,H(t) \frac{\partial \delta_{\mathrm{DM}}}{\partial t} =
   4\pi G  \left[ {\rho}_b \delta_{\mathrm{b}} + {\rho}_{\rm{DM}} \delta_{\rm{DM}} \right]
  \label{eq:deltaDM}
  \end{equation}
Here $a$ is the scale factor, $H = \dot{a}/a$ is the  
Hubble rate, 
$\rho_b$ and $\rho_\gamma$ are respectively the baryon and photon mass density, $n_e$ is the electron number density, $\sigma_T$ is the Thomson cross section for electron-photon scattering,
and $c_b$ is the baryon sound speed.

Prior to recombination, due to the high rate of photon-baryon scatterings, radiation and baryonic matter behave as a sole tightly coupled conducting medium, with the magnetic pressure being insignificant compared to the pressure of the photon-baryon fluid.
As soon as photons start to free stream the evolution of baryonic perturbations can no longer be adequately described by Eq.~\ref{eq:deltab}. 
It is thus necessary to incorporate 
compressional and rotational perturbations in this ionized medium. 
For that the Lorentz force can be formulated using the tangled field, 
\begin{equation}
S(\vec{x}, t) = \frac{\pmb{\nabla}\cdot[\boldsymbol{B}\times(\pmb{\nabla}\times\boldsymbol{B})]}{4\pi{\rho}_b(t_0)}.
\label{equ:S}
\end{equation} 

We can now solve the magnetohydrodynamics equation describing the growth of total matter density perturbations \citep[see, e.g.,][]{1998PhRvL..81.3575S,2005MNRAS.356..778S},

\begin{eqnarray}
      \frac{\partial^2 \delta_m}{\partial t^2}   + 2\,H(t) \frac{\partial \delta_m}{\partial t} = 
      4\pi G \rho_{m}\delta_m + \frac{\rho_b}{\rho_m} \frac{\vec{S}(\vec{x},t)}{a^3},
  \label{eq:deltam}
\end{eqnarray}

where $\delta_m=({\rho}_{\rm{DM}}\delta_{\rm{DM}}+{\rho}_b \delta_b)/\rho_m$, with $\rho_m = (\rho_{\rm{DM}}+\rho_b)$ stands for the summation over both baryonic and dark matter perturbed components.
For scales above the Jeans mass, the term representing fluid pressure, $c_b^2 \nabla^2 \delta_b$, can be neglected.  
Following recombination, the damping term in Eq.~\ref{eq:deltab} attributed to the radiative viscousity can also be neglected, as the damping effect caused by the Hubble expansion becomes more important \citep{jedamzik1998,1998PhRvL..81.3575S}.  
The growth of perturbations generated by any inhomogeneous\footnote{The homogeneous solutions are related to fluctuations produced by sources before recombination.} magnetic field is then found to be

\begin{equation}
    \delta_{m,\rm{PMFs}}(\vec{x},t) \simeq \frac{t^2_{\mathrm{rec}}\,\vec{S}(\vec{x},t_{\mathrm{rec}})}
    {a^3(t_{\mathrm{rec}})}
    \frac{\rho_b}{\rho_m}
    \left[ \frac{9}{10} \left( \frac{t}{t_{\mathrm{rec}}} \right)^{2/3} + \frac{3}{5} \frac{t_{\mathrm{rec}}}{t} -\frac{3}{2}
    \right],
    \label{eq:delta_m}
\end{equation}

where $t_{\rm{rec}}$ is the time of recombination epoch. 
The spatial variation of $\delta_{m,\rm{PMFs}}(\vec{x},t)$ in Eq.~\ref{eq:delta_m} can be tracked through the magnetic source term $S(\vec{x},t)$. Therefore, it is expected that 
the power spectrum of the magnetic field directly affects over-densities on various scales, and ultimately influences the total matter power spectrum.

\begin{figure}
    \centering
    \includegraphics[width=0.48\textwidth]{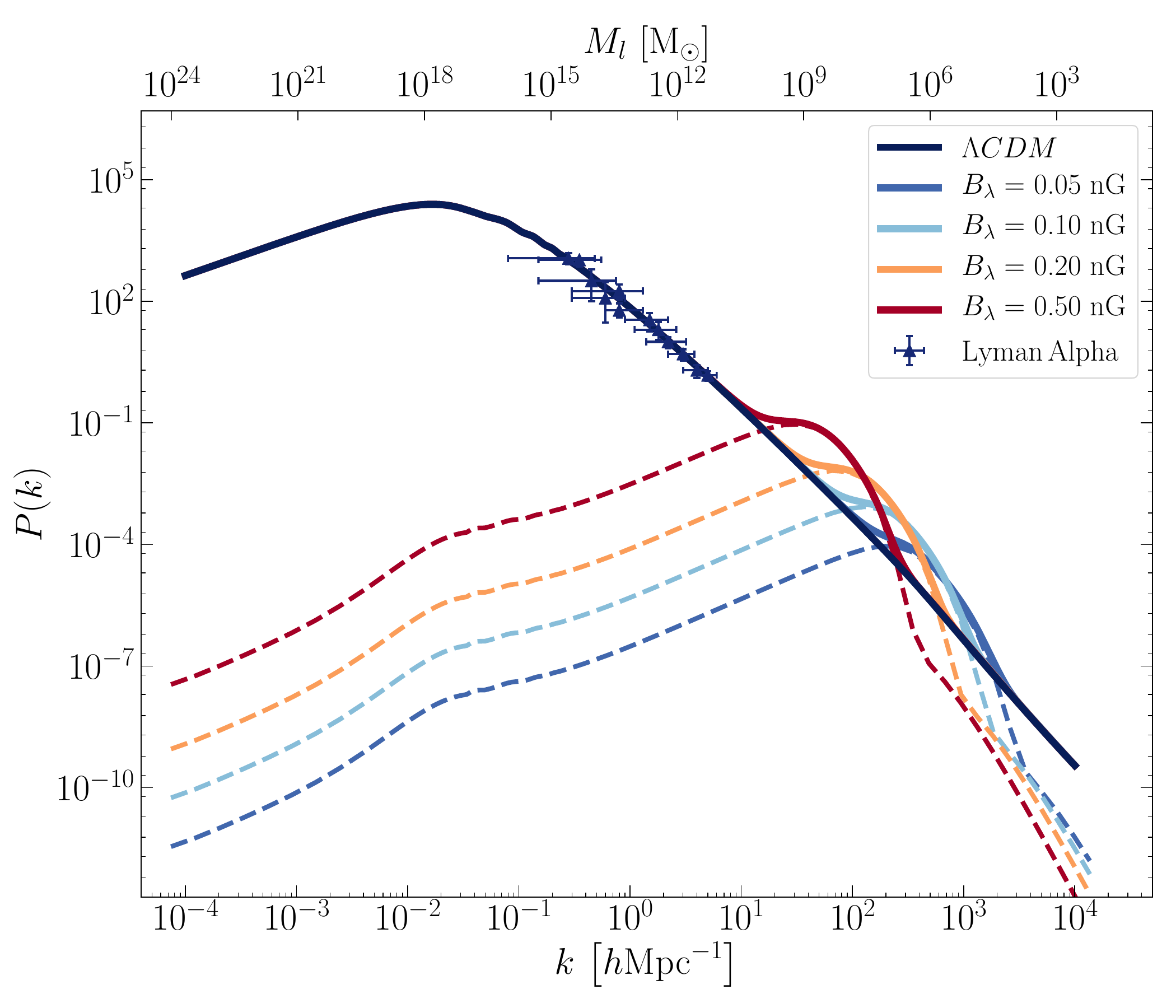}
    \caption{The impact of primordial magnetic fields on the matter power spectrum varying with increasing field strength. Dashed lines illustrate the distribution of matter induced by primordial magnetic fields, while the solid lines represent the combination of these magnetic-induced matter perturbations and the perturbations produced during inflation. The unperturbed $\Lambda$CDM spectrum is shown as the dark blue solid line. The amplitude of the magnetic field is varying from $0.05$ to $0.50\,\mathrm{nG}$, while the spectral slope is kept constant at $n_{B}= -2.9$. The triangles with error bars represent Lyman Alpha data \citep{tegmark2002} as the observational constraints on the matter power spectrum at the smallest mass scales.}
    \label{fig:ps}
\end{figure}
%

\subsection{Impact on the total matter power spectrum}

To understand the effect of primordial magnetic fields on the $\Lambda$CDM matter power spectrum, computing the ensemble average of density perturbations is required. 
In a primordially magnetized early Universe, the total perturbations are the sum of those generated by inflation and the additional perturbations induced by primordial magnetic fields \citep{1996ApJ...468...28K, 2003JApA...24...51G}.
With a Fourier space expression, it leads to
\begin{equation}
    P(k,t) = \langle [ \delta_m(k,t)+\delta_{m,\rm{PMFs}}(k,t)]\,[\delta^*_m(k,t)+\delta^*_{m,\rm{PMFs}}(k,t) ] \rangle.
    \label{eq:pkt}
\end{equation}

%
Figure~\ref{fig:ps} showcases $P(k,t)$ at the present epoch.  
The dark blue line demonstrates the $\Lambda$CDM matter power spectrum, resulting from the ensemble average of $[\delta_m(k,t)]^2$. 
This represents density perturbations obtained from the coupled equations~\ref{eq:deltab} and \ref{eq:deltaDM} in the absence of primordial magnetic fields.
To obtain the matter power spectrum induced by primordial magnetic fields, which is demonstrated by dashed lines, we need the Fourier transform of Eq.~\ref{equ:S}. Because of the two spatial derivatives in $S(x,t)$, this results in $P_{\rm{PMFs}}(k,t) = \langle [\delta_{m,\rm{PMFs}}(k,t)]^2\rangle \propto k^4\,B_{\lambda}^2 \propto A_B\,k^{2n_B+7}$.
At large $k$ modes,
perturbations produced by a scale-invariant magnetic spectrum (i.e. $n_B\cong-3$ and $P_{\rm{PMFs}}(k)\sim k$) dominate the inflationary induced perturbations. 
However, the magnetic pressure leads to an increase of the Jeans length and denotes the lower limit at which perturbations can collapse. 
\citet{1996ApJ...468...28K} suggested a formulation for the magnetic Jeans scale which occurs at the transition between growing modes and oscillatory modes, 
\begin{equation}
    k_{J,B} = \frac{5\pi\rho_b\sqrt{G}}{\vec{B}_{\lambda}}.
    \label{eq:Jeans}
\end{equation}
%
By considering the magnetic Jeans length, we arrive at an expression for the shape of dashed lines in Fig.~\ref{fig:ps}, which corresponds to the matter power spectrum induced by primordial magnetic fields, $\langle [\delta_{m,\rm{PMFs}}(k,t)]^2\rangle$,   \citep{2003JApA...24...51G}, 
\begin{equation}
  P_{\mathrm{PMFs}}(k) \sim A\, k^{2 n_B + 7} + B\, k_{\rm{max}}^{2 n_B + 3} k^4 + C\, k_{\rm{max}}^{2 n_B + 1} k^6 + ...,
\label{eq:pk}  
\end{equation}

Here $A$, $B$ and $C$ are coefficients
which, similarly to $A_B$, 
only depend on the spectral index.
The magnetic counterpart of the Jeans length $k_{\rm{max}}\cong k_{J,B}$ is regarded as the minimum length scale above which the growth of density perturbations is suppressed. 
Equation~\ref{eq:pk} highlights that for $n_B<-1.5$ perturbations on small scales are amplified until they reach $k_{\rm{max}}$, beyond which they are rapidly quenched.
Finally, solid colored lines in Fig.~\ref{fig:ps} indicate the total power spectrum obtained from Eq.~\ref{eq:pkt} with the inclusion of both inflationary and magnetically generated perturbations.

To generate the power spectra presented in Fig.~\ref{fig:ps}, we use a modified version of the \texttt{CAMB} code \citep{2012PhRvD..86d3510S} that incorporates 
primordial magnetic fields.
This version explicitly calculates the non-linear effects of magnetic pressure and the non-negligible viscous damping prior to recombination.
Therefore, there is no need for an artificial cutoff wave number, making the power spectrum more accurate compared to analytical studies in the literature.
The range of parameters explored in Fig.~\ref{fig:ps} is 
carefully selected to affect the power spectrum 
while still adhering to the constraints presented in Section~\ref{sec:intro}.
Therefore, we use a nearly scale-invariant power spectrum with $n_B\ =\ -2.9$, and examine a variety of magnetic field amplitudes, $B_\lambda = 0.05,\ 0.10,\ 0.20,\ 0.50\,\textrm{nG}$. 
On the top axis, we show the link between the scales in Fourier space and the corresponding mass scales in real space. For that, we define the mass enclosed within a comoving Lagrangian sphere of radius $r_l \cong 2\pi/k$ at the present epoch, \citep[see e.g.,][]{2017ARA&A..55..343B},
%
\begin{eqnarray}
M_l &=& \frac{4\pi}{3}r^3_l \rho_m 
= \frac{\Omega_m H^2_0}{2G}r^3_l  \nonumber \\
&=& 1.71\times10^{11} \mathrm{M}_{\odot} \left(\frac{\Omega_m}{0.3}\right) \left(\frac{h}{0.67}\right)^2 \left(\frac{r_l}{1\mathrm{Mpc}}\right)^3, 
\end{eqnarray}
where $H_0$ is the Hubble constant. 
%
With this definition, it becomes evident that the tilt in the matter power spectrum appears in the mass ranges that encompass the dark matter halos of dwarf galaxies. Consequently, we expect that the formation and characteristics of dwarfs could be substantially affected by the presence of primordial magnetic fields.


\section{Numerical methods and simulations}\label{sec:simulations}

We generate all the cosmological 
simulations studied in this work employing our modified version of the \texttt{RAMSES} code \citep{Teyssier2002}.
In addition to the collisionless dark matter and stellar particles,
\texttt{RAMSES} employs an adaptive mesh refinement octree grid to solve the evolution of gas. 
The code models ideal magneto-hydrodynamics \citep[MHD; ][]{2006A&A...457..371F, 2006JCoPh.218...44T} in addition to treating baryonic physics, such as redshift-evolving and uniform UV heating, gas cooling, star formation, and stellar feedback.
Below we provide a brief summary of its essential elements.

\subsection{Numerical setup}\label{sec:setup}

\paragraph{Ideal MHD} 
In its ideal MHD implementation, \texttt{RAMSES} employs a Constrained Transport (CT) method \citep{Teyssier2006,Fromang2006} to solve the equations that govern the evolution of magnetic fields.
The induction equation is solved in a conserved integral form on the cell faces. This requires magnetic fields to be stored as six fields on the cell faces. This is unlike all the hydrodynamic quantities in the simulation, namely, densities, velocities, and energy components which are stored at the center of each gas cell.
The CT method ensures that the magnetic field has zero divergence up to numerical precision, preventing any unwanted modifications of conserved quantities \citep{2000JCoPh.161..605T} or the emergence of magneto-hydrodynamical artifacts \citep{2016MNRAS.455...51H}.

In the ideal MHD setup, assuming a highly conductive medium, the induction equation is
solved with negligible diffusivity but we note that numerical diffusivity is present because finite size of the grid cells. 
Additionally, the distortion of the primordial field due to the velocity perturbations is only significant at sufficiently small scales and can be disregarded at galactic scales \citep[][]{1978PhDT........95W, 1980lssu.book.....P}. 
In the absence of non-ideal magnetic sources such as 
the Biermann battery, 
this leads to ${\partial\, (a^2\vec{B}_{\lambda})}/{\partial t}=0$ 
for the time evolution of the field.   
As the growth of compressional modes is suppressed before recombination, it solves as $\vec{B}_{\lambda}(\vec{x}, t) = \vec{B}_{\lambda}(\vec{x},t_{\mathrm{rec}})\, a^2(t_{\mathrm{rec}})/a^2(t)$, where
$\vec{B}_{\lambda}(\vec{x},t_{\mathrm{rec}})$ refers to the value of the magnetic field at recombination ($t=t_{\mathrm{rec}}$).
In this work, the primordial magnetic field is modeled by an ab-initio $\vec{B}_{\lambda}$ that is seeded uniformly and aligned with the $z$-axis of the computational domain.
This choice of a uniform field is common practice amongst MHD simulations \citep[e.g.,][]{Vazza2014, Pakmor2017, Martin-Alvarez2018}, but we note that smaller-scale perturbations for such strong initial fields may influence the magnetic field amplification and galaxy formation during the early stages of simulations.

\paragraph{Radiative cooling and heating processes} 
In addition to primordial gas cooling, we account for metal-line cooling according to the gas metallicity.
Above temperatures of $10^4\,\mathrm{K}$ we interpolate the corresponding {\sc cloudy} tables \citep{Ferland1998}. Below $10^4\,\mathrm{K}$ we follow fine structure metal cooling rates from \citet{Rosen1995}.
We model the process of reionization, based on the prediction from \citet{Haardt1996}, using a redshift-dependent UV background  which we initialize at $z =10$. 
Hydrogen self-shielding against the ionizing radiation is incorporated by suppressing the UV-background heating for gas densities above $n_{\mathrm{H}}=0.01\,\mathrm{g}\,\mathrm{cm}^{-3}$.
To account for the pre-enrichment by the Population III stars we assume a metallicity floor of $\left[ \mathrm{Fe}/\mathrm{H} \right] = -5$. 
Above this metallicity, fine-structure line cooling of atomic carbon and
oxygen \citep{Bromm2003}, lead to gas fragmentation and the formation of low-mass stars.

\paragraph{Star formation}
We model the process of star formation employing a magneto-thermo-turbulent (MTT) star formation prescription, presented in more detail in the hydrodynamical version by \citet{Kimm2017} and \citet{Trebitsch2017}, and in its MHD version by \citet{2020MNRAS.495.4475M}. 
As a brief summary of this star formation model, gas is allowed to transform into stellar particles only in cells that are at the highest level of refinement, where the local combination of magnetic, thermal and turbulent support is overcomed by the gravitational pull \citep{Rasera2006}.   
The conversion of gas into stars in star forming cells follows the Schmidt law \citep{Schmidt1959},
\begin{equation}
    {\dot\rho_{\star}} = \epsilon_{\mathrm{ff}}\frac{\rho_{\mathrm{gas}}}{t_{\mathrm{ff}}}.
\end{equation}
Here the star-forming gas with a density of $\rho_{\mathrm{gas}}$ is converted into stars, with free-fall time $t_{\mathrm{ff}}$. 
In our formulation, the efficiency of this conversion $\epsilon_{\mathrm{ff}}$ is not a constant, but rather computed locally based on the gas properties of each region. 
It is calculated following 
the multi-scale model of \citep{2011ApJ...730...40P} as presented in \citet{Federrath2012}. 

\paragraph{Stellar feedback}

Each stellar particle in our simulations corresponds to a single stellar population that is characterized by its own initial mass function (IMF).
The IMF is modeled as a probability distribution function following a Kroupa shape \citep{Kroupa2001} and normalized over the complete range of masses.
This allows each stellar particle of mass $294.81\,\mathrm{M}_{\odot}$ to stochastically populate stars within the mass interval of $\left[0.05 - 50\right]\,\mathrm{M}_{\odot}$ during the initial $50\,\mathrm{Myr}$ of its formation.  
The number of exploding supernovae for each particle is then calculated at each time step based on the lifetimes of stars it contained. 

We use the mechanical supernovae feedback prescription of \citet{2014ApJ...788..121K}, for stellar particles undergoing a supernova event.
In this approach, the momentum injected by supernova explosions is determined by the physical characteristics of the gas being swept up, such as its density $n_{\mathrm{H}}$, and metallicity $Z$, 
\begin{equation}
    p_{\mathrm{SN}}(E,\, n_{\mathrm{H}},\, Z) \approx 3\times 10^{5}\,\mathrm{km}\,\mathrm{s}^{-1}\,\mathrm{M}_{\odot}\,E_{51}^{16/17}\,n_{\mathrm{H}}^{-2/17}\,f(Z),
\end{equation}
where the momentum input is decreasing with metallicity as $f(Z) =\mathrm{max\left[Z/Z_{\odot}, 0.01\right]^{-0.14}}$. Here $E_{51}$ is the energy in the unit of $10^{51}\,\mathrm{erg}$, and $n_{\mathrm{H}}$ the hydrogen number density.
This method ensures that the feedback from supernovae is accurately modeled at all blast wave stages, from the initial free expansion to the final momentum-conserving snowplow phase.
Along with the momentum, energy is deposited into the neighboring cells. 
The specific energy of each supernova (SN) has a value 
$\varepsilon_\text{SN} = E_\text{SN} / M_\text{SN}$, where 
 $E_\text{SN} = 10^{51} \mathrm{erg}$ and $M_\text{SN} =10\, \mathrm{M}_{\odot}$.
Each supernova
also returns a fraction of stellar mass back to the ISM.
We use $\eta_\text{SN} = 0.213$ for fraction of $M_\text{SN}$ returned as gas mass, and $\eta_\text{metal} = 0.075$ for the 
newly synthesized metals. 
This set of values is selected to match the model galaxies in the fiducial setting with the observed global properties of Local Group dwarfs (See Sec.~\ref{sec:scaling_relations}).
Our feedback prescription does not include stellar winds, local radiation from young stars, photo-electric heating, and cosmic rays, all of which, in addition to supernovae feedback, can potentially reduce the final stellar mass of our galaxies.
%

\begin{table*}[t]
\caption{\small
  Parameters varying in the simulation runs. 
  Columns are as follows: 
  1) Model ID.  
  2) Comoving strength of the initial primordial magnetic field. 
  3) Matter power spectrum used to generate the initial conditions.
  \label{tab:params}}
  \centering
\begin{tabular}{l c c c c c |c c c c}
 \hline
 \hline
 \texttt{ID} & \texttt{B00} & \texttt{B05} & \texttt{B10} & \texttt{B20} & \texttt{B50} & 
 \texttt{B05ps} & \texttt{B10ps} & \texttt{B20ps} & \texttt{B50ps} \\

  $B_{\lambda}\,[\mathrm{nG}]$ & 0 & 0.05 & 0.10 & 0.20 & 0.50 & 0.05 & 0.10 & 0.20 & 0.50 \\ 

 Power Spectrum & \multicolumn{5}{c|}{\texttt{$\Lambda$CDM}} & \multicolumn{4}{c}{\texttt{$\Lambda$CDM+PMFs}} \\
  \hline
  \hline

\end{tabular}
\end{table*}

\subsection{Initial conditions}
The initial conditions are generated using the \texttt{MUSIC} code \citep{2011MNRAS.415.2101H},
and the cosmology of \citet{Planck2016} with $\Omega_\Lambda = 0.685$,  
$\Omega_m = 0.315$, $\Omega_b = 0.0486$, and $h = 0.673$.
All simulations are started at redshift $z = 200$, ensuring that the rms variance of the initial density field, $\sigma_{8}$, lies between 0.1 and 0.2 \citep{2009ApJ...698..266K,onorbe2015}. 
%
Except for a small subset, all of our simulations are run until redshift $z = 0$.

The twelve halos studied in this work are selected from different regions in the dark matter only cosmological box of \citet{Revaz2018}.
Those regions are identified to form a dark matter halo with a mass typical of spheroidal and ultra-faint dwarf galaxies at redshift $z = 0$. 
We re-simulate those halos, 
including the full treatment of baryons described above.
Using the zoom-in technique, for each halo we refine a 3D ellipsoid of size about $0.85\,\mathrm{cMpc}$ 
across, positioned in the center of the cubic simulation box with $L_{\mathrm{Box}}=5.11\,\mathrm{cMpc}$ per side.
The size of the ellipsoid is determined such that it encompasses all particles that eventually reside within each target halo by redshift $z=0$ \citep{onorbe2014}.
In this refined region, we achieve dark matter mass resolution of 
$m_{\mathrm{DM}}\simeq4\cdot10^{3}\,\mathrm{M}_{\odot}$.
We gradually degrade the resolution, from level 10 to 6 of \texttt{MUSIC} outside the zoom region \footnote{One resolution level $l$ corresponds to $N=(2^l)^3$ particles in the full cosmological box. The particle mass is thus decreased by a factor of eight between two levels.}.

Initially, the domain is discretized with a uniform grid of $1024^3$ cells. This resolution is preserved within the zoom region, while the grid is de-refined to level 6 elsewhere. 
Throughout the course of the simulation, the adaptive refinement criteria come into play to effectively resolve dense and Jeans-unstable regions.
When the total dark matter and gas mass within a grid cell exceeds $8\,m_{\mathrm{DM}}$, or when the size of the grid cell surpasses $4$ local Jeans length, a parent grid cell is split into $8$ equal child cells. 
This process follows the octree structure of \texttt{RAMSES}, where
the size of cell $i$ is determined by the refinement level of cell $l_i$ according to $\Delta x_{i} = 1/2^{l_i}\,L_{\mathrm{Box}}$. 
In our simulations, with a maximum refinement level of $20$, the initial grid undergoes adaptive refinement to achieve a minimum cell width of approximately $5\,\mathrm{pc}$.

These initial conditions and the configuration of the simulations focus on capturing and resolving the formation of galaxies in halos within the perturbed range of the total matter power spectrum in Fig.~\ref{fig:ps}.

\begin{table*}[h]
  \caption{\small Global properties of the eight halos simulated in our fiducial hydrodynamical model \texttt{B00}, at redshift $z=0$. The first column gives the halo ID following \citet{Revaz2018}. $L_{\rm{V}}$ is the V-band luminosity and $M_{\star}$ the stellar mass. $M_{\rm{200}}$ is the virial mass, i.e., the mass inside the virial radius $R_{\rm{200}}$. $\sigma_{\rm{LOS}}$ is the line-of-sight velocity dispersion and [Fe/H] is the abundance ratio of iron with respect to hydrogen. 
    \label{tab:catalogue}}
  \centering
  \resizebox{0.65\textwidth}{!}{%
  \begin{tabular}{l c c c c c c c }
  \hline
  \hline

    Halo ID & $L_{\rm{V}}$           & $M_{\star}$            & $M_{\rm{200}}$          & $R_{\rm{200}}$  &  $r_{1/2}$ & $\sigma_{\rm{LOS}}$ & [Fe/H] \\
       & $[10^5\,\rm{L_\odot}]$ & $[10^5\,\rm{M_\odot}]$ & $[10^9\,\rm{M_\odot}]$  & $[\rm{kpc}]$  & $[\rm{kpc}]$  & $[\rm{km/s}]$ & [dex] \\ 
  \hline
  \hline
\texttt{h025} & 9.55 & 19.67 & 8.90 & 65.6 & 0.57 & 9.4 & -1.95 \\
\texttt{h070} & 1.02 & 2.42 & 2.16 & 40.2 & 0.50 & 6.0 & -2.73 \\
\texttt{h063} & 0.58 & 1.44 & 2.49 & 42.2 & 0.82 & 4.4 & -2.41 \\
\texttt{h177} & 0.31 & 0.78 & 0.61 & 26.4 & 0.45 & 4.5 & -2.42 \\
\texttt{h111} & 0.16 & 0.38 & 1.25 & 33.6 & 0.35 & 5.0 & -2.94 \\
\texttt{h159} & 0.24 & 0.59 & 0.78 & 28.7 & 0.42 & 4.8 & -2.69 \\
\texttt{h277} & 0.09 & 0.21 & 0.90 & 30.0 & 2.00 & 5.9 & -2.65 \\
\texttt{h190} & 0.06 & 0.15 & 0.57 & 25.8 & 0.72 & 3.8 & -3.04 \\
  \hline
  \hline
  \end{tabular}
  }

\end{table*}
\subsection{Set of simulations}

We run nine sets of simulations as listed in Table~\ref{tab:params}. 
The run which has no magnetic fields and employs 
a $\Lambda$CDM initial configuration
is our reference, hereafter denoted as \texttt{B00}.
%
In the first series of simulations, 
a uniform magnetic field is introduced, aligned with the $z$-axis of the simulation box. 
The field has an initial comoving strength varied from $0.05$ to $0.50\,\mathrm{nG}$, which later evolves self-consistently throughout the simulation.    
The initial matter power spectrum is set using the $\Lambda$CDM paradigm, where no additional perturbation due to primordial magnetic fields is included.  
Models in this set are denoted by \texttt{B05}, \texttt{B10}, \texttt{B20}, \texttt{B50}.   
%
In the second series of simulations, in addition to the 
initial uniform field of comoving strength ranging 
from $0.05$ to $0.50\,\mathrm{nG}$, the effect of primordial fields on the matter power spectrum is taken into account. In these models, denoted as \texttt{B05ps}, \texttt{B10ps}, \texttt{B20ps}, \texttt{B50ps}, the initial conditions are generated from a \texttt{$\Lambda$CDM+PMFs} power spectrum. In each model, the amplitude of the primordial magnetic field is the same as the uniform field in the MHD configuration initiated at redshift $z=200$, and the spectral index is set to $n_{B}=-2.9$ to have a scale-invariant magnetic power spectrum.

All the simulations are designed to investigate primordial magnetic fields with strengths allowed by current cosmological constraints (see Section~\ref{sec:intro}).
From Fig.~\ref{fig:ps} it appears that a stronger amplitude than the maximum one used in this work ($B_{\lambda}=0.50\,\mathrm{nG}$) would
affect the power spectrum at scales of $k<10\,h\,\rm{Mpc^{-1}}$. However, such a modification is ruled out by the observations of the Lyman-$\alpha$ forest \citep{tegmark2002,2013ApJ...762...15P}.
The smallest amplitude employed ($B_{\lambda}=0.05\,\mathrm{nG}$) lies at the threshold where primordial magnetic fields start to affect the $\Lambda$CDM power spectrum and potentially exert an influence on structure formation \citep{2010PhRvD..81d3517S, pandey2015, sanati2020}.
%

\subsection{Extraction of the observables, luminosity and metallicity}\label{sec:obs}
\paragraph{Galactic region}
Global galactic properties, such as gas mass and specific energies, are measured inside the \textit{galactic region}.
It is defined using the virial radius of the dark matter halo $R_{200}$ at redshift $z=0$. 
The galactic region is the spherical volume centered on the position of the galaxy with radius equal to the comoving evolution of $\sim1.5\,R_{200}\, (z = 0)$. 
This method is employed to adaptively increase the size of the studied region as the galaxy and its hosting dark matter halo grow in size. 
In order to obtain the centre of the galaxy, we tag the system by finding its position at redshift $z \sim 7$.
From that point onwards, at each coarse time-step of {\sc ramses} we re-compute the position 
of the galaxy centre. 
To do this, we simply obtain the new center of mass of the most central $\sim 50$ stellar particles in the previous iteration. 
We use this approach to compute various galaxy properties on-the-fly, with time resolution equivalent to that of the coarse time-stepping of the simulation.

\paragraph{V-band luminosity} 
The galaxy V-band luminosity ($L_V$) is obtained by summing the luminosities of all stellar particles located within $1.8$ times the halo's half-light radius. Their mass is converted into luminosity using the stellar population synthesis model of \citet{vazdekis1996} computed with the revised \citet{Kroupa2001} IMF. 
Where necessary, the luminosities are inter- and extra-polated in age and metallicity using a bivariate spline.
We ignore dust absorption, as its role only becomes critical for intrinsic magnitudes brighter than $M_{AB}\sim-20$ \citep{10.1093/mnras/sty1024}.

\paragraph{Metallicity}
Due to the limited number of stars forming in dwarf galaxies, selecting a representative \textit{mean} $[\rm{Fe/H}]$ is a non-trivial undertaking.
Since the metallicity distribution is sparsely sampled, calculating the mode (peak) of the metallicity distribution function can result in significant uncertainties.
In this study, we use the method proposed in \citet{2023A&A...669A..94S} to determine the mode based on a fitting analytical formula derived from a simple chemical evolution model. The error on  $[\rm{Fe/H}]$ is taken at the maximum of the errors obtained by the different methods, peak and mode. 

\paragraph{Velocity dispersion} 
The line-of-sight stellar velocity dispersion, $\sigma_{LOS}$, is calculated for seven different lines of sight inside a $1\,\rm{kpc}$ cylindrical radius. The value quoted for each galaxy represents the mean of these values.


\section{Results}\label{sec:results}

Our simulations follow the evolution of six different ultra-faints and two spheroidal dwarf galaxy with final luminosities $\sim 10^3 - 10^5\,{L_\sun}$.
Table~\ref{tab:catalogue} summarizes the main properties of model galaxies in our fiducial hydrodynamical model \texttt{B00}, at redshift $z=0$. 
IDs are those of \citet[][see their Table 1]{Revaz2018}. The seven other columns show the galaxy total V-band luminosity $L_{\mathrm{V}}$, total stellar mas $M_{\star}$, virial mass $M_{200}$, virial radius $R_{200}$, mean stellar velocity dispersion $\sigma_{\rm LOS}$ and peak value of the metallicity distribution function $[\rm{Fe/H}]$.

\subsection{Impact of magnetic fields on gas distribution}\label{sec:dens}

\begin{figure*}
    \centering
    \includegraphics[width=\textwidth]{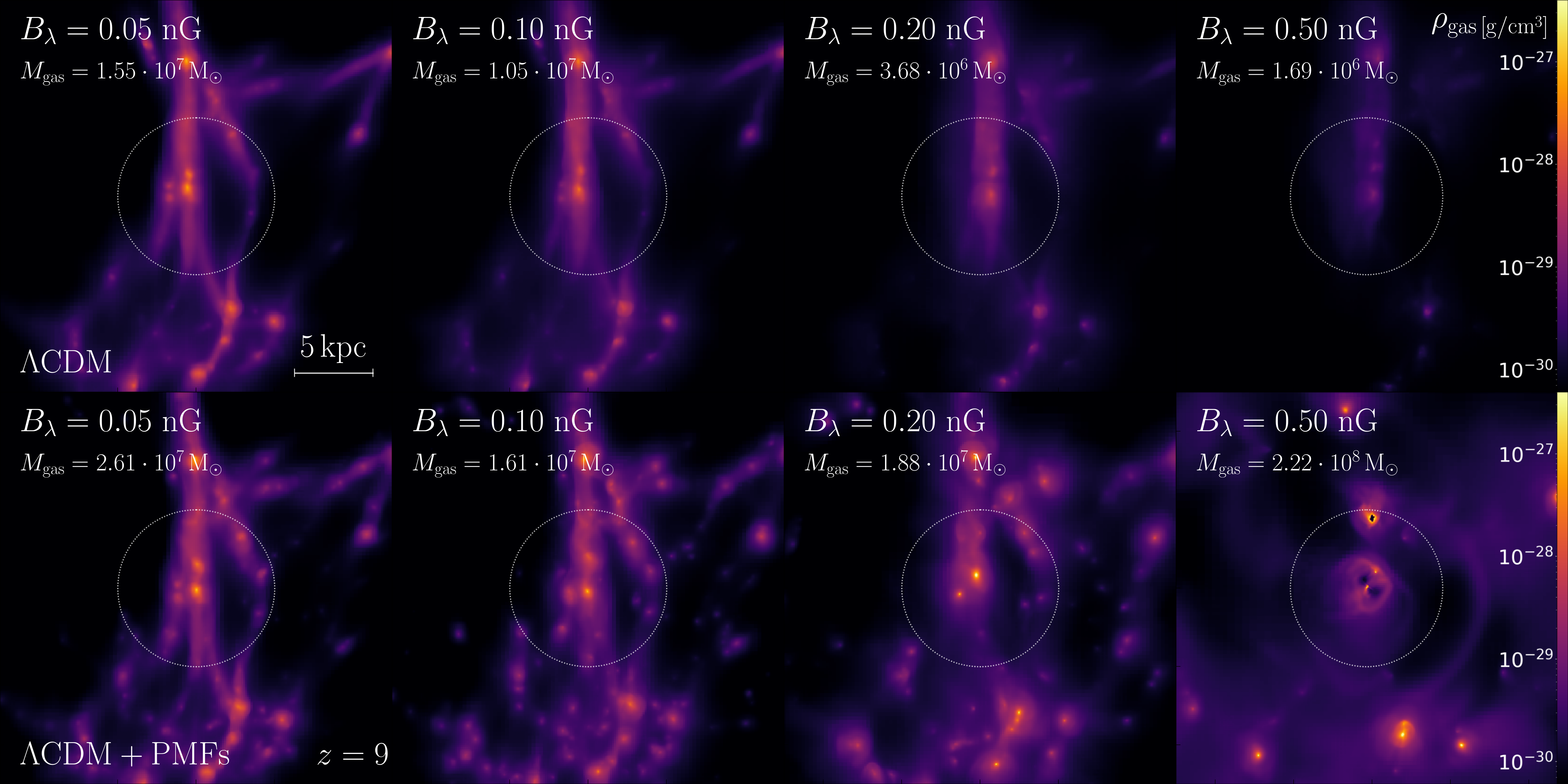}
    \caption{Gas density projections of model galaxy \texttt{h070}. 
    The strength of the magnetic field is increasing from $B_{\lambda}=0.05\,\mathrm{nG}$ on the left to $B_{\lambda}=0.50\,\mathrm{nG}$ at the rightmost panels. The upper and lower rows represent all the models generated without primordial magnetic fields (\texttt{$\Lambda$CDM}), and with primordial magnetic fields (\texttt{$\Lambda$CDM+PMFs}) matter power spectrum, respectively. 
    Snapshots are taken at redshift $z=9$.
    The white circle shows the galactic region, 
    which is traced in all time steps. The evolution of various galaxy quantities, such as the gas content, is evaluated inside this region with radius $\sim1.5$ virial radius of the dark matter halo. Panels have physical sides and depth of $256\,\mathrm{kpc}$. The included gas mass represents the average mass of gas in the galactic region 
    over the first $2\,\mathrm{Gyr}$.}
    \label{fig:gasDensity1}
\end{figure*}

Figure~\ref{fig:gasDensity1} shows the density-weighted projection of gas density in 
the simulated galaxy \texttt{h070} (see Table~\ref{tab:catalogue}). 
The magnetic field strength is increasing from $0.05\,\mathrm{nG}$ in the leftmost to $0.50\,\mathrm{nG}$ in the rightmost panel. 
Top and bottom panels correspond to \texttt{$\Lambda$CDM} and \texttt{$\Lambda$CDM+PMFs} simulation setup, respectively.  
Snapshots
are taken at redshift $z=9$, which corresponds to 
the intense phase of star formation 
in dwarf galaxies \citep{okamoto2012,brown2014, sacchi2021, gallart2021}. 
The snapshot in model \texttt{B50ps} is shown at redshift $z=13$ where it stopped due to computational expenses. 

We compute the gas mass in each snapshot 
inside the galactic region (see Section~\ref{sec:obs}) which is illustrated by the white circles. The galactic region covers the main halo of the targeted dwarf galaxy and its surrounding in a $1.5\times\,\mathrm{R}_\mathrm{vir}$ radius. $M_{\mathrm{gas}}$ in the top left of each panel, shows the gas mass within the galactic region averaged over the first $2\,\mathrm{Gyr}$, before the dwarf is completely depleted of gas.

The upper panels correspond to the first set of simulations, where the magnetic fields are 
initialized as uniform values, 
as described in Section~\ref{sec:simulations}, and the matter power spectrum is that of the classical $\Lambda$CDM paradigm. 
These simulations are designed to explore the contribution of magnetic fields in shaping the internal structure 
of dwarf galaxies.
Here, magnetic fields generate additional pressure support. 
Along with other hydrodynamical pressures such as the thermal or kinematic, this counteracts the gravitational field and effectively hampers the collapse of gas.
It particularly affects the first stages of galaxy formation, i.e., for $z>6$. 
The result is that the gas collapse in the first minihalos is delayed by 
several hundred $\mathrm{Myrs}$.
The gravitational field in minihalos is therefore less likely to retain the gas when it is heated and pushed out by the UV-radiation at lower redshifts. 
When $B_{\mathrm{\lambda}} = 0.50\,\mathrm{nG}$ the collapse of gas is completely halted, as shown in the most right panel. 
This effect leads to a reduction trend in the averaged gas mass with increasing $B_{\lambda}$.
%

\paragraph{}
The bottom panels of Fig.~\ref{fig:gasDensity1} represent the second set of simulations, where in addition to varying the strength of the magnetic field seeds, we include their effect on the
matter power spectrum. 
These simulations are designed to explore the net effect of primordial magnetic fields 
through the altered matter power spectrum on the one hand, and the direct magnetically-produced turbulence and pressure, on the other hand. 
Here the amplitude of the primordial fields varies from $B_{\lambda}=0.05$ to $0.50\,\mathrm{nG}$. 
The first two columns correspond to models with relatively weaker magnetism.
The gas density maps in these two models resemble their counterparts in the first row. 
Indeed, the impact of weak primordial fields on the matter power spectrum and consequently the buildup of galaxies is slim \citep[see also][]{2005MNRAS.356..778S, sanati2020, 2021MNRAS.507.1254K}. 
However, the average gas mass experiences a slight increase, which can be attributed to the additional perturbations induced by primordial magnetic fields in the matter distribution. These perturbations lead to an earlier formation of the first dark matter minihalos and
an earlier onset (approximately $100\,\mathrm{Myrs}$) of gas collapse, which subsequently continues at a higher rate.

Increasing $B_{\lambda}$ introduces considerable reformations in the main halo of the galaxy and the number of its subhalos. 
Dark matter only simulations of \citet{sanati2020} shows a strong increase in the maximum of the Gaussian halo mass function, up to a factor of 7, when $B_{\lambda}=0.20$ and 
$0.50\,\mathrm{nG}$, compared to their model without magnetically perturbed initial density fluctuations.
For $B_{\lambda}=0.20\,\mathrm{nG}$, 
Fig.~\ref{fig:ps} suggests that the additional perturbations generated by primordial magnetic fields mainly affect halos with the Jeans mass $M_{l}\leq10^8\,\mathrm{M_{\odot}}$. This impact is reflected in Fig.~\ref{fig:gasDensity1}, where the number of subhalos in this mass range is noticeably increased. Moreover, $M_{\mathrm{gas}}$ inside the galactic region, which contains the main halo of the targeted dwarf and its subhalos, is increased.

When $B_{\lambda}=0.50\,\mathrm{nG}$, 
from Fig.~\ref{fig:ps} it appears that the altered matter power spectrum influences the formation of halos with mass $M_{l}\simeq10^9\,\mathrm{M_{\odot}}$.
Therefore, in this model, the main halo of the dwarf galaxy is primarily affected. 
In the final panel of Fig.~\ref{fig:gasDensity1} it is evident that the gas mass is approximately $100$ times greater compared to the counterpart derived from the \texttt{$\Lambda$CDM} paradigm depicted in the first row.
Here, the strong gravitational potential of the central halo, 
can merge the satellite subhalos and leads to the formation of a massive dwarf galaxy.

\subsection{The evolution of energy components}\label{sec:energy}
\begin{figure*}
    \centering
    \includegraphics[width=0.85\textwidth]{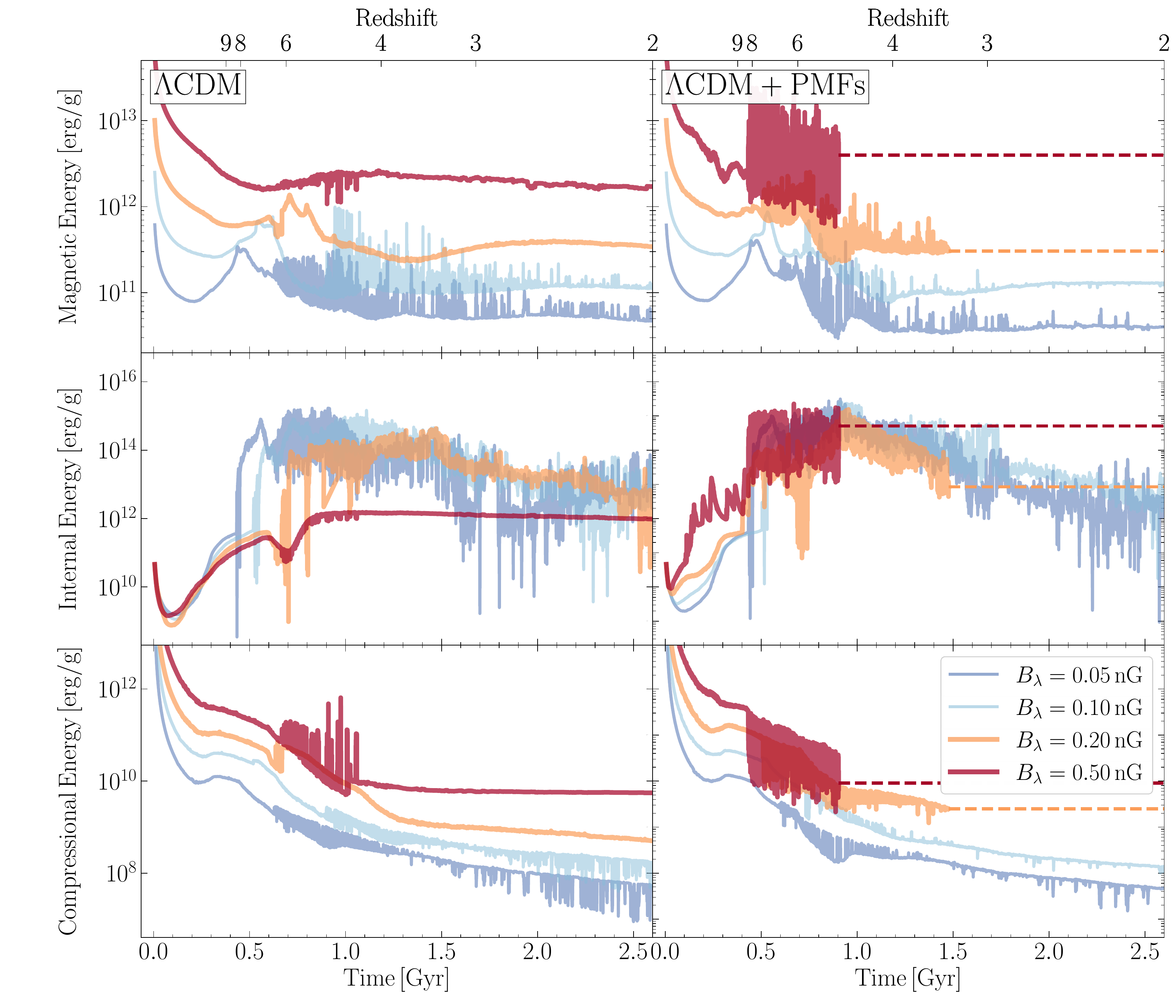}
    \caption{Comparison of different specific energy components within the galactic region (white circle in Fig.~\ref{fig:gasDensity1}) of \texttt{h070}. The left and right panels represent all the MHD runs emerging from a standard \texttt{$\Lambda$CDM} and \texttt{$\Lambda$CDM+PMFs} matter power spectrum, respectively. 
    The upper panels show the time evolution of specific magnetic energy. The ratio of magnetic energy between different models with initial seeds between $0.05$ to $0.50\,\mathrm{nG}$ increases by a factor smaller than $B_{\lambda}^2/8\pi$, suggesting energy loss due to magnetic field saturation, including energy dissipation during the collapse. The middle panels show the specific internal energy, which tracks the onset and rate of star formation. The lower panels show the estimate for the compressional part of the magnetic energy from isotropic adiabatic gas collapse. Magnetic energy initially follows the compressional energy as $B_{\lambda}^2 \propto \rho_{\mathrm{gas}}^{4/3}$, until magnetic field amplification causes a deviation in their time evolution. 
    } 
    \label{fig:energy}
\end{figure*}
In this section, we explore the co-evolution of the key components shaping the energy allocation as a dwarf galaxy evolves over time.
The three panels in Fig.~\ref{fig:energy},
respectively from top to bottom, represent 
the time evolution of the magnetic energy, internal energy and our estimate for the compressional evolution of the magnetic energy. 
For this plot we use dwarf spheroidal \texttt{h070}, but the outcomes are similar for other halos examined in this work (see Table~\ref{tab:catalogue}).   
Note that the gas content within the galaxy halo, at different time steps, is subject to the 
natural mass growth due to
accretions, mergers and outflows.
Equally, computing the total energy over time is sensitive to the shape and volume of a galaxy.   
Therefore, for a fair comparison between different energy components, we use 
specific energies
\begin{eqnarray}
\varepsilon_{x} = \frac{E_{x}}{M_{\mathrm{gas}}},
\end{eqnarray}
as the ratio between the total energy $E_{x}$ of component $x$
and the gas mass $M_{\mathrm{gas}}$, inside the galactic region (see Section~\ref{sec:obs}).
The main amplification of the magnetic energy is expected to occur before $z=2$ \citep{2008Natur.454..302B}. During this period, gas kinetic energy is converted into magnetic through magnetic field amplification by a small-scale dynamo, reaching equipartition levels 
\citep{2003PASA...20..252G}.
After this redshift, magnetism is observed to be in equipartition with other energy components \citep{1992ApJ...388...17W,2013MNRAS.433.1675B,2015A&ARv..24....4B}.  
For that reason and for clarity, we truncate the time evolution of energies in this plot at $2\,\mathrm{Gyr}$.
Each line represents a different model from Table~\ref{tab:params}, where 
color codes are the same as in Fig.~\ref{fig:ps}. 
In the left column, the galaxy simulated using a classical matter power spectrum with the strength of the magnetic field varying from $B_{\lambda}=0.05$ to $0.50\,\mathrm{nG}$, showcases \texttt{B05}, \texttt{B10}, \texttt{B20}, and \texttt{B50} models. 
In the right column, the magnetically-induced matter perturbations by primordial fields are also included in the models.

\paragraph{Magnetic energy}
The topmost panels show the specific magnetic energy $\varepsilon_{\mathrm{mag}}$ as it evolves over time.
%
The time evolution of the magnetic field is described by the induction equation:
\begin{equation}
    \frac{1}{a}\frac{\partial{(a^2\vec{B}_{\lambda})}}{\partial t} = \vec{\nabla} \times \left[ \vec{v} \times(a^2\vec{B}_{\lambda}) - \frac{1}{a}\eta\,\vec{\nabla} \times (a^2\vec{B}_{\lambda}) \right]. 
\end{equation}
In the absence of resistivity ($\eta=0$), peculiar velocities ($\vec{v}=0$) and non-ideal magnetic sources such as Biermann battery \citep{1950ZNatA...5...65B}, 
this leads to ${\partial\, (a^2\vec{B}_{\lambda})}/{\partial t}=0$.  
Before density fluctuations start collapsing into structures, both magnetic fields and gas density are decreasing with the expansion of the Universe as $1/a^2$ and $1/a^3$, respectively, shaping the relation $B_{\lambda}^2 \propto \rho_{\mathrm{gas}}^{4/3}$.
The magnetic flux freezing to the plasma 
is imprinted in the early decaying of the specific magnetic energy at high redshifts, as shown in the topmost panels of Fig.~\ref{fig:energy}.

As expected, models with stronger initial seeds ($B_{\lambda}=0.20$ and $0.50\,\mathrm{nG}$)
reach higher magnetic energy values
during their evolution.
This pattern is evident in both the unperturbed $\Lambda$CDM set of simulations and the magnetically-perturbed set, as depicted in the left and right columns, respectively.

It is interesting that the energy ratios in each set of runs do not increase by a factor of $B_{\lambda}^2/8\pi$. For instance, when we compare the \texttt{B0.05} and \texttt{B0.10} models, one might anticipate a $2^2$ increase in energy due to a stronger initial magnetic seed. However, a smaller energy increase of $\sim 2.7$ suggests 
energy loss due to magnetic field saturation, 
with some energy dissipation even occurring during the collapse phase. The saturation of magnetic energy becomes more pronounced in stronger models and can occur within relatively short time intervals, such as a few hundred $\mathrm{Myrs}$.
This behaviour is followed in both sets of simulations presented in the left and right panels. 
In models featuring weaker magnetization ($B_{\lambda}=0.05$ and $0.10\,\mathrm{nG}$), 
it takes a longer period for the magnetic energy to 
reach saturation.

\paragraph{Internal energy}
The middle panels show the specific internal energy $\varepsilon_{\mathrm{int}}$, as a function of time.
In the \texttt{$\Lambda$CDM} setup, the internal energy gradually decreases as the value of $B_{\lambda}$ increases from $0.05$ to $0.50\,\mathrm{nG}$. 
This decline can be attributed to the supplementary pressure support exerted by magnetic fields,
which also opposes the collapse of gas clouds (see the discussion in Section~\ref{sec:dens}) and hinders star formation.
Given that the internal energy closely tracks the heating generated by the stellar feedback, raising the magnetic field strength leads to a decrease in
the number of peaks and valleys, as well as the amplitude of
$\varepsilon_{\mathrm{int}}$.

This reduction in internal energy is particularly pronounced during the early stages of galaxy formation, specially at redshifts $z>6$, distinguishing the stronger models \texttt{B0.20} and \texttt{B0.50} from the weaker models \texttt{B0.05} and \texttt{B0.10}.   
In the extreme case of \texttt{B0.50}, the star formation is dramatically suppressed and a limited increase of  $\varepsilon_{\mathrm{int}}$ is observed.
Another notable characteristic evident here is the delay in the growth of internal energy when the magnetic field strength is significant, particularly $B_{\lambda}=0.20$ and $0.50\,\mathrm{nG}$. 
The evolution of internal energy closely corresponds to changes in density and temperature and consequently reflects the collapse of gas clouds. Therefore, the delay caused by strong magnetic fields in the formation of minihalos impacts the timing at which $\varepsilon_{\mathrm{int}}$ begins to accumulate from its initial value.

\paragraph{}
In the \texttt{$\Lambda$CDM+PMFs} setup,
a distinct trend is evident in the right panel compared to the left panel, showcasing the influence of primordial fields. 
Here, a higher variability is evident in the internal energy, characterized by numerous peaks and valleys, specially in stronger models. This pattern, visible across all strengths of magnetic field, implies a more dynamic star formation history.  
In these models, the formation of more massive minihalos accelerates the pace of collapse, star formation and consequently influences the rate at which internal energy is growing.
%
Additionally, primordial magnetic fields expedite the onset of minihalo formation.
As a result, the collapse of gas clouds is accelerated, leading to the onset of internal energy accumulation within timescales as short as several tens of $\mathrm{Myr}$, specifically when $B_{\lambda}=0.50\,\mathrm{nG}$.

\paragraph{Compressional energy}
The bottom panels show the evolution of the specific compressional energy $\varepsilon_{\mathrm{comp}}$ versus time. This term is obtained from an isotropic adiabatic approximation of the gas collapse. 
During the early stages of galaxy formation, 
the magnetic energy density scales with the gas density as $\propto \rho_{\mathrm{gas}}^{4/3}$ \citep[see e.g.,][]{2017MNRAS.472.4368R, Martin-Alvarez2018}. 
Thus $E_{\mathrm{comp}}(t_0)= E_{\mathrm{mag}}(t_0)\propto{\rho_{\mathrm{gas}}^{4/3}}(t_0)\times\mathrm{volume}(t_0)$.
In the absence of amplification or decaying processes, this approximation dictates the evolution of the frozen-in magnetic field lines, which closely follow the collapse of cosmic structures, indicating $E_{\mathrm{comp}}(t)\propto{\rho_{\mathrm{gas}}^{4/3}}(t)\times\mathrm{volume}(t)$.
For extracting this term in each time step of the simulation, we use
\begin{eqnarray}
E_{\mathrm{comp,cell}}(t)= 
\frac{dx^3_{\mathrm{cell}}}{\frac{4}{3}\pi r^3_{\mathrm{gal}}(t_0)} \times
\langle E_{\mathrm{mag,gal}}(t_{0}) \rangle \times
\left( \frac{\rho_{\mathrm{gas,cell}}(t)}{\langle\rho_{\mathrm{gas,gal}}(t_0)\rangle}\right)^{4/3}, 
\end{eqnarray}
%
where the compressional energy in each cell of size $dx_{\mathrm{cell}}$ and at each time step $t$, denoted as $E_{\mathrm{comp,cell}}(t)$, is directly proportional to the gas density $\rho_{\mathrm{gas,cell}}(t)$ within that particular cell.   
%
$\langle E_{\mathrm{mag,gal}}(t_0) \rangle$ stands for the average of magnetic energy within the galactic region of radius $r_{\mathrm{gal}}$ at time $t_0$ when it is at its minimum state.  
This corresponds to the turnaround point, marking the moment when the initial perturbations transition from expansion to collapse. 
At this moment, when the magnetic energy reaches its lowest value, baryonic matter accumulates and the first stars begin to form within the emerging galaxy.
Eventually, the total compressional term is derived by summing over all cells within the galactic region. It is then divided by the total gas mass in this region to obtain the specific energy.  
By comparing the evolution of magnetic energy and compressional energy, we can investigate the nature of magnetic field amplification. 
In case of dynamo amplification, it is anticipated that magnetic energy undergoes exponential growth, and the ratio of magnetic energy to compressional energy significantly exceeds $1$.

\begin{figure}
    \centering
    \includegraphics[width=0.48\textwidth]{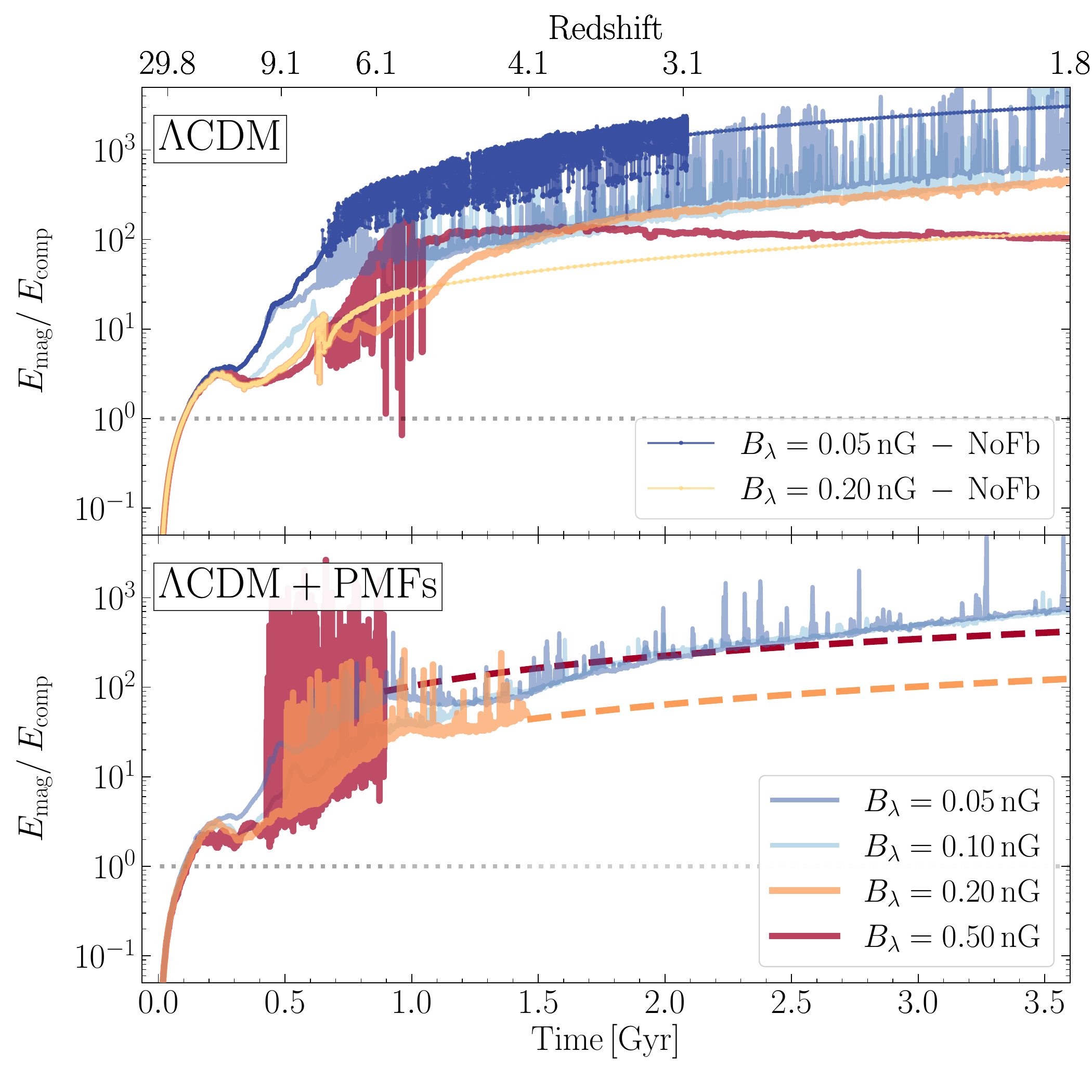}
    \caption{Time evolution of the ratio of magnetic energy to compression energy in the galactic region (white circle in Fig.~\ref{fig:gasDensity1}) for all the simulations during the first $3.5\,\mathrm{Gyr}$. The upper and lower panels corresponds to all the models in \texttt{$\Lambda$CDM} and \texttt{$\Lambda$CDM+PMFs} settings, respectively. The solid and dotted lines correspond to the runs with the fiducial value of supernovae feedback 
    $E_{\rm{SN}}=10^{51}\,\mathrm{erg}$ and zero feedback, 
    respectively.}
    \label{fig:dynamo}
\end{figure}

Figure~\ref{fig:dynamo} presents the ratio between $E_{\mathrm{mag}}$ and $E_{\mathrm{comp}}$ as a function of time. As described, the compressional term demonstrates the expected buildup of magnetic fields, when exclusively frozen in the collapse of matter into dark matter halos. 
The deviation of $E_{\mathrm{mag}}$ from $E_{\mathrm{comp}}$, apparent in all models, 
indicates the amplification of magnetic fields through different mechanisms that occur alongside adiabatic magnetic compression. 
These mechanisms include the 
activity of a dynamo 
within collapsing gas clouds. 
In both upper and lower panels, the ratio initially follows the same curve in all models, as magnetic energy evolves with the expansion of the Universe as $B_{\lambda}\propto a^{-2}$,
until 
the density perturbations start to collapse. Once the galaxy reaches the turnaround point from expansion to collapse in each model, the ratio takes different paths. Increasing the magnetic field strength results in a slower amplification rate. This is due to a considerably higher $B_{\lambda}(t_0)$ and $B_{\lambda}$ in the galaxy. Thus it takes significantly longer time for the stronger models to reach a similar amplification rate as the weaker ones. The upper and lower panel showcase the same result in \texttt{$\Lambda$CDM} and \texttt{$\Lambda$CDM+PMFs} setting.

It is worth noting that this amplification can also be attributed to the influence of powerful supernova feedback events.
In this scenario, the supernova shock evacuates the surroundings of the exploding star from gas, thus reducing the $E_{\mathrm{comp}}$. 
Simultaneously, the supernova winds push the magnetic fields outwards, while keeping them in the galactic region.  
Therefore the $E_{\mathrm{mag}}/E_{\mathrm{comp}}$ ratio is increasing with time.

%
In order to unravel the source of separation between the compressional and magnetic energy terms, 
we re-simulate \texttt{h070} in model \texttt{B05} and \texttt{B20}, deactivating the supernovae feedback.
The comparison is shown in the upper panel of 
Fig.~\ref{fig:dynamo}.
The dotted lines represent the simulations in which the supernovae feedback is set to zero.
The more rapid increase in magnetic energy compared to compressional energy remains evident in simulations without feedback. 
This indicates that the amplification of magnetic fields is not solely attributed to the supernova explosion process. 
It further strengthens the notion that magnetic fields initiated with even a week seed, e.g., $B_{\lambda}=0.05\,\mathrm{nG}$ can undergo dynamo amplification as the galaxy evolves through gas accretion, halo collapse, and mergers.
The higher ratio of $E_{\mathrm{mag}}/E_{\mathrm{comp}}$ in models without supernovae feedback is because magnetic field back reactions from supernovae explosions slow down the growth of magnetic energy in models with feedback.

\subsection{Impact of magnetic fields on star formation}\label{sec:sfr}
%

In this section, we present the influence of primordial magnetic fields on the star formation history and final stellar mass of dwarf galaxies.
From each zoom-in simulation, we first identify the halo corresponding to the reference halo in the fiducial model. Subsequently, using the position and particle IDs, we extract the correspondent halo in other models. The physical quantities of each galaxy such as the stellar mass are computed inside the virial radius $R_{200}$ of the extracted halo. 

\begin{figure}
  \centering
    \includegraphics[width=0.48\textwidth]{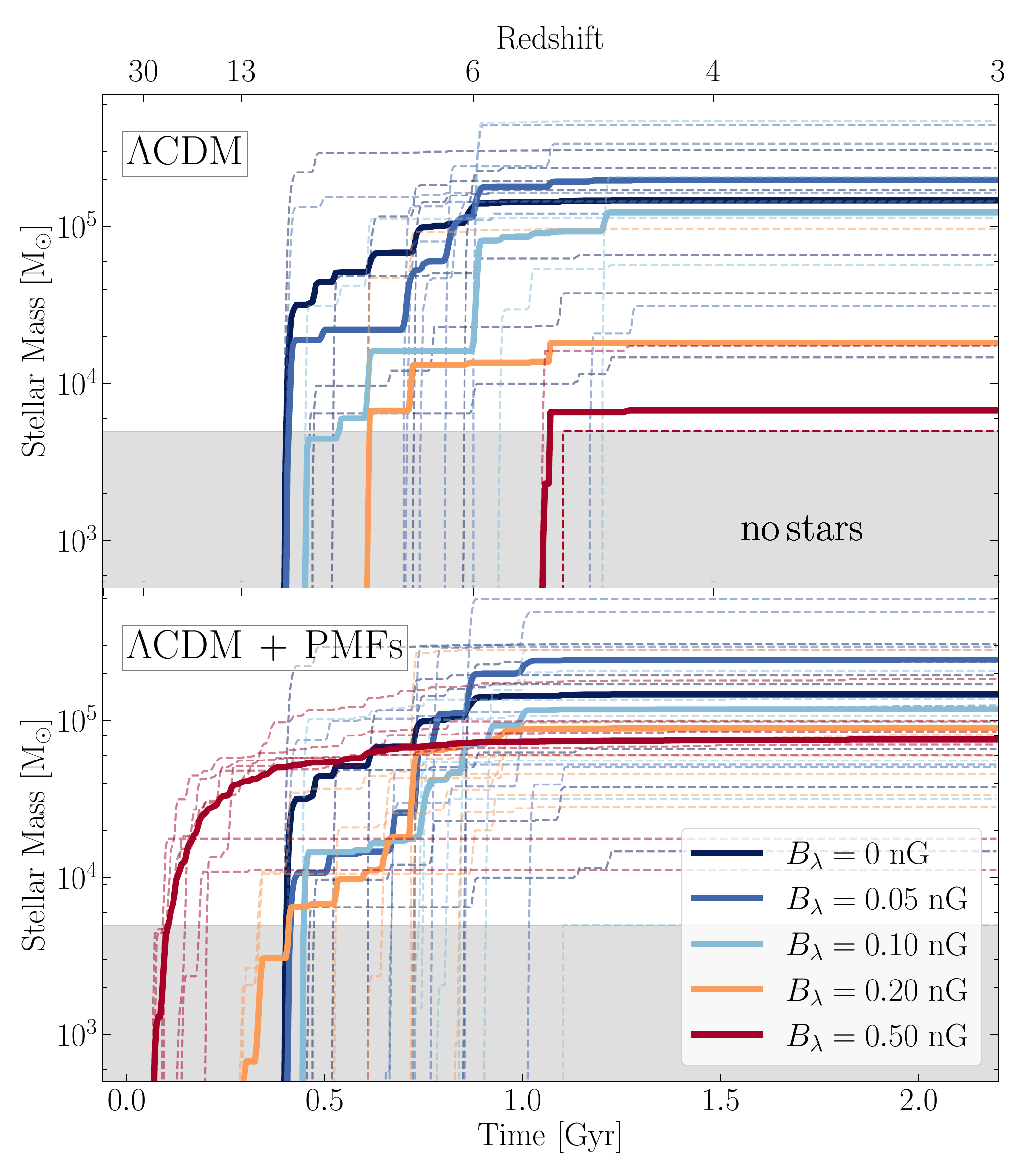}
    \caption{Cumulative stellar mass during the first $2\,\rm{Gyr}$. 
    Each dashed line represents the stellar mass of each model galaxy simulated in a magnetic field model with strength varying from $B_{\lambda} = 0.05$ to $0.50\,\rm{nG}$, compared with the model without magnetic field. Each solid line represents the mean stellar mass of all galaxies simulated in each model. 
    The upper and lower panels 
    correspond
    to all the models in \texttt{$\Lambda$CDM} and \texttt{$\Lambda$CDM+PMFs} settings, respectively. Top) Magnetic fields delay the formation of first stars. In low mass halos of ultra-faint dwarf galaxies, $B_{\lambda} = 0.50\,\rm{nG}$ completely prevents the star formation. In order to include halos which do not form any stars in the plot, their stellar masses are shown at $4\cdot10^{3}\,M_{\odot}$. Bottom) The influence of primordial magnetic fields on the matter power spectrum accelerates the formation of the first dark matter halos, leading to an earlier onset of star formation but an overall reduction in the final stellar mass.}
    \label{fig:SF_ps}

\end{figure}

Figure~\ref{fig:SF_ps} displays the cumulative mass of stars forming within the initial $2\,\rm{Gyr}$. In each magnetic field model, the line represents the average value of stellar mass measured for all galaxies simulated using the same magnetic field strength. 
In the fiducial model \texttt{B00}, shown in dark blue, star formation is quenched in the simulated galaxies before $1\,\rm{Gyr}$ at the latest. 

In the first set of simulations, 
the delay induced in the onset of star formation in the MHD configuration can be seen in the top panel of Fig.~\ref{fig:SF_ps} across all amplitudes of magnetic field examined. This effect is specifically evident in the first $\mathrm{Gyr}$. 
%
Following the delay in the birth of first stars, it is difficult for galaxies to catch up with the cumulative star formation in the fiducial model, when the gas is heated by the background UV-radiation.
This delay becomes more apparent in models with strong magnetic fields, $B_{\lambda}=0.20$ and $0.50\,\mathrm{nG}$, which have a more substantial influence 
on the gas distribution.
%
In weaker models, $B_{\lambda}=0.05$ and $0.10\,\mathrm{nG}$, this delay results in a shortened star formation history of less than $500\,\mathrm{Myr}$, particularly pronounced in ultra-faint dwarf galaxies with a small gas mass.
The rate of star formation and the final stellar mass are consistently decreasing as the strength of the magnetic field increases.


%

The bottom panel of Fig.~\ref{fig:SF_ps} shows models which include magnetic fields through the altered matter power spectrum \texttt{$\Lambda$CDM+PMFs}.   
One striking outcome of these models is an expedited onset of star formation.  
This stems from the influence 
of primordial fields in speeding up the formation of first dark matter halos and their subsequent gas accretion. 
This effect persists across all amplitudes 
of magnetic fields examined,
further confirming the results obtained in \citet{sanati2020}, for different amplitudes and spectral indices of primordial magnetic fields. 
However, it is more moderate in models with a weak primordial field.
From Fig.~\ref{fig:ps} it can be seen that the region affected by weakly perturbed models (\texttt{B0.05ps} and \texttt{B0.10ps}) encompasses halos with a Jeans mass of approximately $10^5-10^7\,\mathrm{M_{\odot}}$. 
Consequently, the weakly perturbed models exhibit a slight increase in the abundance of small minihalos within this mass range. 
However, the gas content in most of these minihalos is insufficient to form stars. 
In these models, the accelerated build up of the galaxies 
is therefore a secondary effect of primordial magnetic fields when considering stellar masses.  Their dominant effect is further counteracting pressure in the ISM.

Increasing the amplitude of primordial magnetic fields affects the matter power spectrum at higher halo masses, as shown in Fig.~\ref{fig:ps}.
%
When $B_{\lambda}=0.20\,\mathrm{nG}$, the abundance of subhalos with a Jeans mass of approximately $10^8\,\mathrm{M_{\odot}}$ increases. 
The gas collapse in these halos is able to satisfy the MTT prescription for the formation of stellar particles. The star formation rate, specially during the first $500\,{\rm Myr}$, is therefore considerably boosted in this model. 
Similarly, when $B_{\lambda}=0.50\,\mathrm{nG}$, the main halo of each dwarf galaxy grows in mass. 
Unlike lower amplitudes of $B_{\lambda}$, for these two models the impact of primordial magnetic fields in accelerating galaxy evolution outweighs their contribution to the overall pressure support and ISM physics.
Figure~\ref{fig:SF_ps} shows that in these two strongly perturbed models star formation is initiated earlier and at higher redshifts compared to the fiducial \texttt{B00} model. This implies that primordial magnetic fields can be a potential mechanism to explain the efficient and rapid star formation observed in galaxies at $z\sim8.5-14.5$ in the CEERS survey \citep[e.g.,][]{2023arXiv231104279F, 2024MNRAS.527.11372}.

%
\begin{figure*}[]
    \centering

    \includegraphics[width=0.85\textwidth]{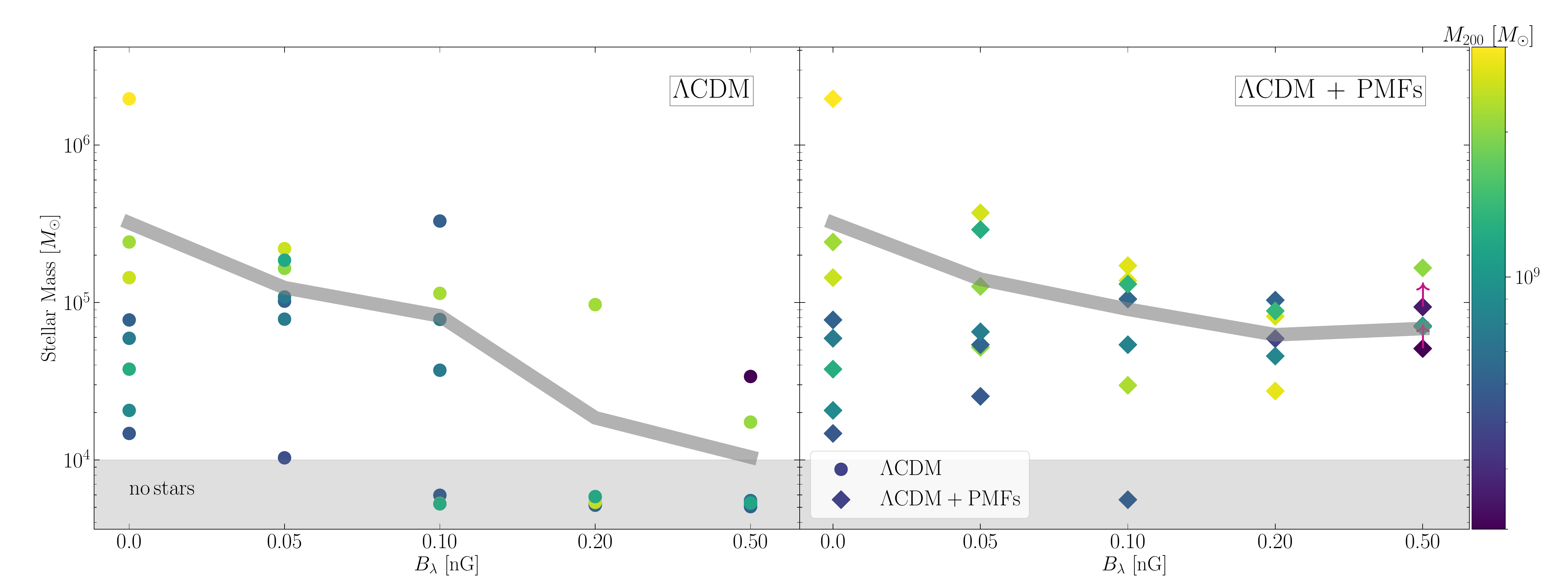}
    \caption{Stellar mass of each galaxy at redshift $z=0$ for varying primordial magnetic field strength, color-coded by the virial mass of the dark matter halo. In the left panel, each circle represents one galaxy simulated in the MHD suite (\texttt{$\Lambda$CDM} model). In the right panel, galaxies shown as empty diamonds (\texttt{$\Lambda$CDM+PMFs} model) consider both magneto-hydrodynamics and the contribution of primordial fields to the matter power spectrum. The stellar mass of halos which do not form any stars is 
    shown at $M_\star < 10^{4}\,M_{\odot}$ instead of $0\,M_{\odot}$.
    Halos that do not reach redshift $z=0$, due to extensive CPU-hours required , are identified by marking them with an upward arrow. 
    }
    \label{fig:MsB}
\end{figure*}
%

\paragraph{}

To quantitatively demonstrate the correlation between star formation and magnetic field strength, Fig.~\ref{fig:MsB} presents the final stellar mass of dwarf galaxies as a function of $B_{\lambda}$. 
In the left panel, each circle is representative for one galaxy at redshift $z=0$, color-coded by the virial mass of its dark matter halo. 
The gray line represents the average stellar mass for all simulated galaxies as a function of magnetic field strength.
For both ultra-faints and dwarf spheroidal, the final stellar mass remains largely unaffected in the weak models with $B_{\lambda} = 0.05\,\mathrm{nG}$. 
However, comparing the dwarf spheroidal to ultra faint dwarf galaxies in stronger magnetic field models shows that the low-mass gas content of ultra faints is very sensitive to any perturbations, namely, the impact of magnetic fields. 
Although the merging history and the star formation history in each ultra-faint galaxy is different, the average trend shows a steady decrease in the final stellar mass.  

In the \texttt{B0.20} model, 
only the dwarf spheroidal galaxy within our sample has a non-zero final stellar mass, while the collapse of minihalos and star formation in all of the simulated ultra-faint dwarfs is completely hindered by a magnetic field with a strength of $B_{\lambda}=0.20\,\mathrm{nG}$.
For illustrating these 
halos in Fig.~\ref{fig:MsB}, their zero stellar mass is replaced by random values between $5$ and $6\times10^3\mathrm{M}_{\odot}$, and marked by \textit{no stars} sign. 
A stronger magnetic field with a strength of $B_{\lambda}=0.50\,\mathrm{nG}$ prevents the formation of stellar particles with our MTT star formation model 
in all the ultra-faint dwarf galaxies.
Here, the most massive dwarf spheroidal in our sample, \texttt{h025},  is used to obtain a threshold for the mass of dwarf galaxies where the impact of magnetic fields become more important in their evolution.
Due to computational constraints, simulations for this galaxy are limited to the fiducial model \texttt{B0.00} and the extreme MHD model \texttt{B0.50}.
By comparing this more massive dwarf with lower mass ultra-fain galaxies, it appears that magnetic fields severely quench the star formation in dwarf galaxies with halo mass and stellar mass below $M_{200}\sim 2.5\cdot10^9\,M_{\odot}$ and $M_{\star}\sim 3\cdot10^6\,M_{\odot}$, respectively.
%

The right panel of Fig.~\ref{fig:MsB} shows models which include the impact of primordial magnetic fields on the matter power spectrum (\texttt{$\Lambda$CDM+PMFs}) in the MHD setup.
Notwithstanding the modifications occurring in the internal structure of dwarf galaxies at high redshifts, the impact of 
magnetic fields in the ISM remains a vital factor in their evolution throughout the star-forming phase. The result of this dual effect on the final stellar mass of model galaxies is displayed in colored empty diamonds.
In strongly perturbed models, \texttt{B20ps} and \texttt{B50ps}, despite an increased number of minihalos with a larger gas content, the persistent effect of additional magnetic energy in the ISM counteracting gravitational collapse throughout the entire star formation period leads to a lower stellar mass compared to the fiducial \texttt{B00} model at redshift $z=0$. 
The noteworthy observation here is that the final stellar mass in halos emerging from the second set of simulations (\texttt{$\Lambda$CDM+PMFs}) surpasses that of their counterparts in the first set (\texttt{$\Lambda$CDM}). Therefore, for dwarf galaxies with luminosity $L_{\mathrm{V}}<10^5\,L_{\odot}$, solely incorporating magnetic fields in numerical simulations while disregarding the influence of primordial fields on their host dark matter halos may result in miscalculation of the impact of primordial magnetic fields in reducing the star formation rate, leading to an overestimation for stronger magnetic field strengths.

\subsection{Impact of magnetic fields on scaling relations 
}\label{sec:scaling_relations}

In this section, we explore the impact of primordial magnetic fields on the global properties of dwarf galaxies. Namely, the V-band stellar luminosity, 
stellar mass, 
virial mass, 
line-of-sight velocity dispersion, 
half-light radius, and stellar metallicity. 
The method for calculating each quantity is described in Section~\ref{sec:obs}.  

\subsubsection{Stellar mass-halo mass}\label{sec:M200Ms}
%
\begin{figure}
    \centering
    \includegraphics[width=0.48\textwidth]{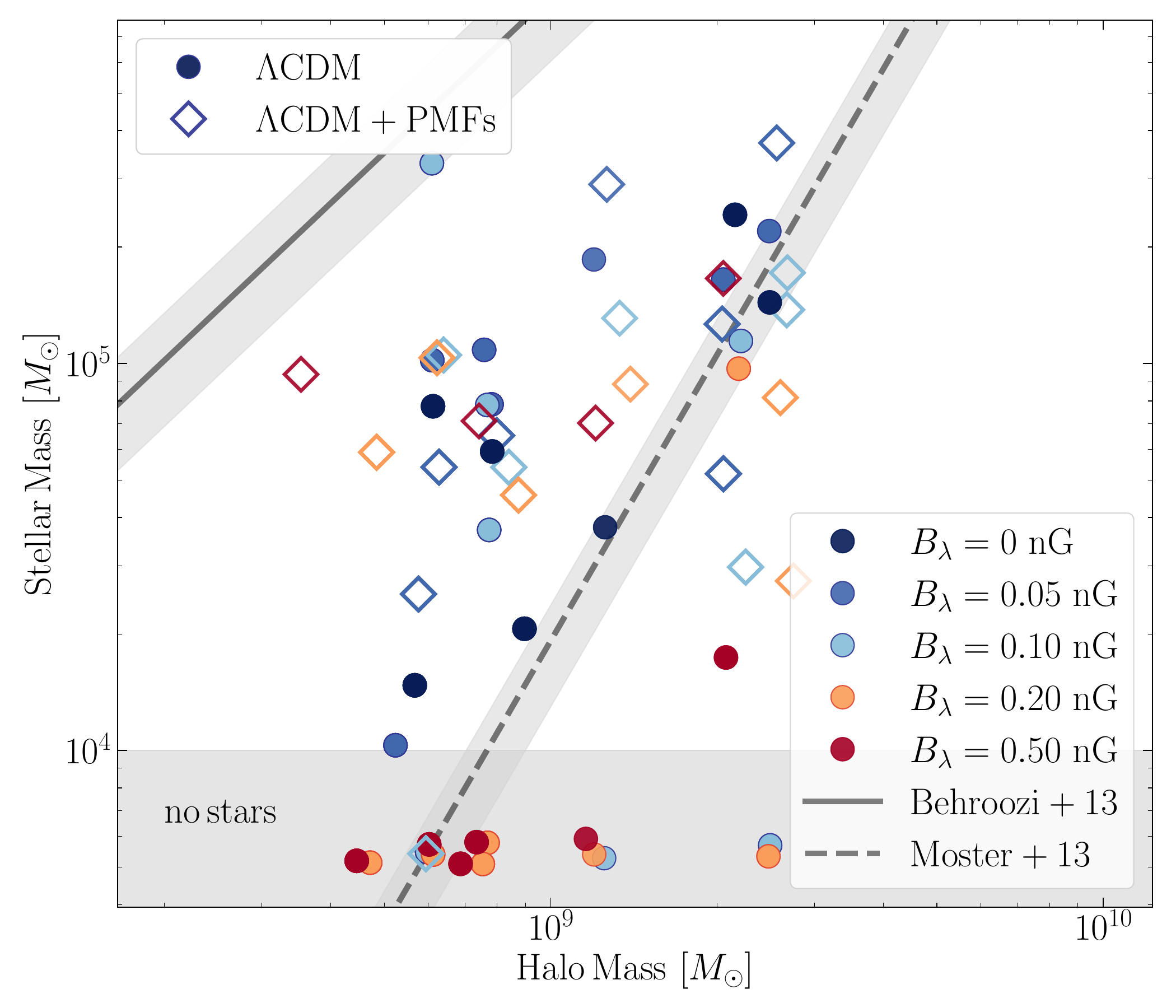}
    \caption{
    Relation between galaxy stellar mass and dark matter halo mass for each simulated dwarf galaxy at redshift $z=0$. For comparison, the extrapolated abundance matching models of \cite{2013ApJ...770...57B} and \cite{2013MNRAS.428.3121M} are shown in solid and dashed lines, respectively. In order to include halos which do not form any stars in the plot, their stellar masses are shown at $M_\star \sim 
    10^{3}\,M_{\odot}$.}
    \label{fig:M200}
\end{figure}
%
Figure~\ref{fig:M200} shows the relation between galaxy stellar mass and dark matter halo mass for each simulated dwarf galaxy at redshift $z=0$.  
The results from the abundance matching models of \citet{2013ApJ...770...57B} and \citet{2013MNRAS.428.3121M} are represented by solid and dashed lines, respectively. Both models have their halo masses extrapolated to 
$10^8\,\mathrm{M}_{\odot}$ to encompass the mass range of the simulated ultra-faint dwarf galaxies. 
%
In the fiducial model \texttt{B00}, the resulting stellar masses range from approximately $10^4$ to $2.5\cdot10^5\,\mathrm{M}_{\odot}$. All galaxies in this model lie well within the expected values for the final halo mass, which spans from $8\cdot10^8$ to $2.5\cdot10^9\,\mathrm{M}_{\odot}$.
Note that the stellar masses of these galaxies are lower than that those of the galaxies simulated in \citet{2018A&A...616A..96R} and \citet{sanati2020}, despite having the same halo mass. This is due to the star formation and stellar feedback schemes varying in different hydrodynamical codes.
%

Colored circles show halos in the first set of simulations, originating from a $\Lambda$CDM paradigm.
Magnetism of low strength ($B_{\lambda} = 0.05$ and $0.10\,\mathrm{nG}$) does not dramatically influence their location in the $M_{\star}-M_{200}$ relation compared to the fiducial model. 
%
Nonetheless, 
the inverse scaling of the stellar mass with the strength of the field, observed in Section~\ref{sec:sfr}, 
persists in this relation as well. 
For a given halo mass, the final stellar mass decreases with increasing $B_{\lambda}$. 
Low mass halos which do not form stars in the strong magnetic field model, where $B_{\lambda} = 0.20$ and $0.50\,\mathrm{nG}$, 
are labeled with \textit{no stars}. To include these halos in the plot, their zero stellar mass is substituted with a random value of $5-6\times10^3\mathrm{M}_{\odot}$. 
%

Colored empty diamonds depict halos in the second set of simulations, emerging from a modified $\Lambda$CDM paradigm (\texttt{$\Lambda$CDM+PMFs}). 
The global properties of galaxies 
undergo only minor changes in the weakly perturbed models. 
When $B_{\lambda} = 0.05$ and $0.10\,\mathrm{nG}$, all halos closely resemble their counterparts in the fiducial model. 
However, in the strongly perturbed models, the final stellar mass recurrently reveals that the influence of primordial magnetic fields on the initial phases of galaxy formation, primarily through modifications to the matter power spectrum, outweighs its subsequent effects in the ISM.
Galaxies which do not form stars in the \texttt{B0.20} and \texttt{B0.50} model, evolve into bright dwarfs with average $M_{\star}=6$ and $7\cdot10^4\,M_{\odot}$ in the \texttt{B0.20ps} and \texttt{B0.50ps} models, respectively. 
Since the abundance matching relation is not well constrained in the regime of ultra-faint dwarf galaxies, none of the models is ruled out based on this comparison.
It is indeed likely that the same halo mass supports orders of magnitude differences in stellar mass, owing to various environmental effects 
and unique history of each dwarf galaxy \citep[see i.e., ][]{2018A&A...616A..96R}.
This may lead to a considerable scatter
in the low mass regime.  

\subsubsection{Luminosity-velocity dispersion}\label{sec:sigma}
%
\begin{figure}
    \centering
    \includegraphics[width=0.48\textwidth]{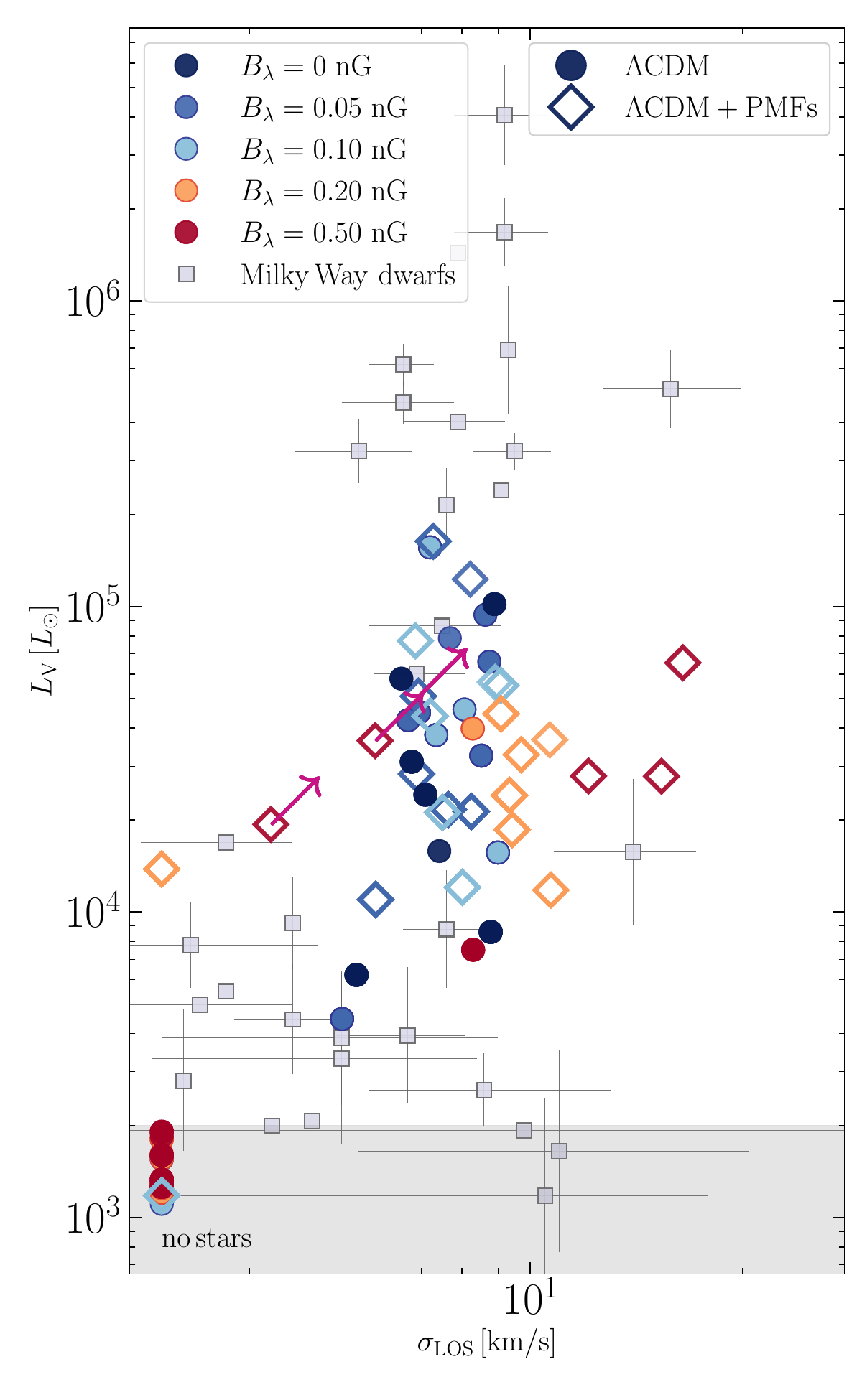}and 
    \caption{V-band luminosity as a function of line-of-sight velocity dispersion in each model galaxy is compared to the observational data of Milky Way satellites in gray squares. In the \texttt{$\Lambda$CDM} setting, increasing the magnetic field strength results in fainter systems (illustrated by colored circles) and can even completely prevent star formation in ultra-faint dwarf galaxies. In order to include halos which do not form any stars in the plot, their luminosities are represented at $L_{\rm{V}} \sim 10^{3}\,\rm{L_{\odot}}$. In the \texttt{$\Lambda$CDM+PMFs} setting, despite lower luminosities, velocity dispersion of galaxies (illustrated by colored empty diamonds) increases due to a higher frequency of mergers.
    }
    \label{fig:sigma}
\end{figure}
%
Figure~\ref{fig:sigma} shows the V-band luminosity $L_{\mathrm{V}}$ versus the line-of-sight velocity dispersion $\sigma_{\rm LOS}$ for each simulated galaxy at redshift $z=0$. 
The comparison to the observational data is obtained from the regularly updated Local Group and Nearby Dwarf Galaxies database of \citet{mcconnachie2012}, represented by gray squares in the figures. 

In the first set of simulations (\texttt{$\Lambda$CDM}), represented by colored circles,
higher values of $B_{\lambda}$ result in a reduced final stellar mass for a given halo mass.
Therefore, increasing the strength of magnetic fields, causes the systems to become noticeably fainter than their corresponding analogues in the fiducial model, or even completely stops the star formation.
Additionally, magnetic fields impact the line-of-sight velocity dispersion in each halo. 
In galaxies where magnetic fields of strength $B_{\lambda}=0.05\,\mathrm{nG}$ lead to less concentrated star formation and an increased half light radius (see Sec.~\ref{sec:r12}), there is an associated increase of the velocity dispersion.

In the second set of simulations (\texttt{$\Lambda$CDM+PMFs}), represented by colored empty diamonds, incorporating primordial magnetic fields through the matter power spectrum, 
both the luminosity and the stellar velocity dispersion of the simulated galaxies are modified.
Particularly, in the strongly perturbed models, \texttt{B20ps} and \texttt{B50ps}, despite lower luminosities, the line-of-sight velocity dispersion of each halo is notably larger than in the fiducial model. 
In these models, due to the impact of primordial magnetic fields on the evolution of dark matter substructures, stars in the main halo of each galaxy migrate from numerous minihalos. In addition, because of greater number of minihalos forming in these models, there is increased frequency of mergers between minihalos.
This results in the observed differences in the dynamics of the simulated galaxies.

\subsubsection{Size-Luminosity}\label{sec:r12}
%
\begin{figure}
    \centering
    \includegraphics[width=0.48\textwidth]{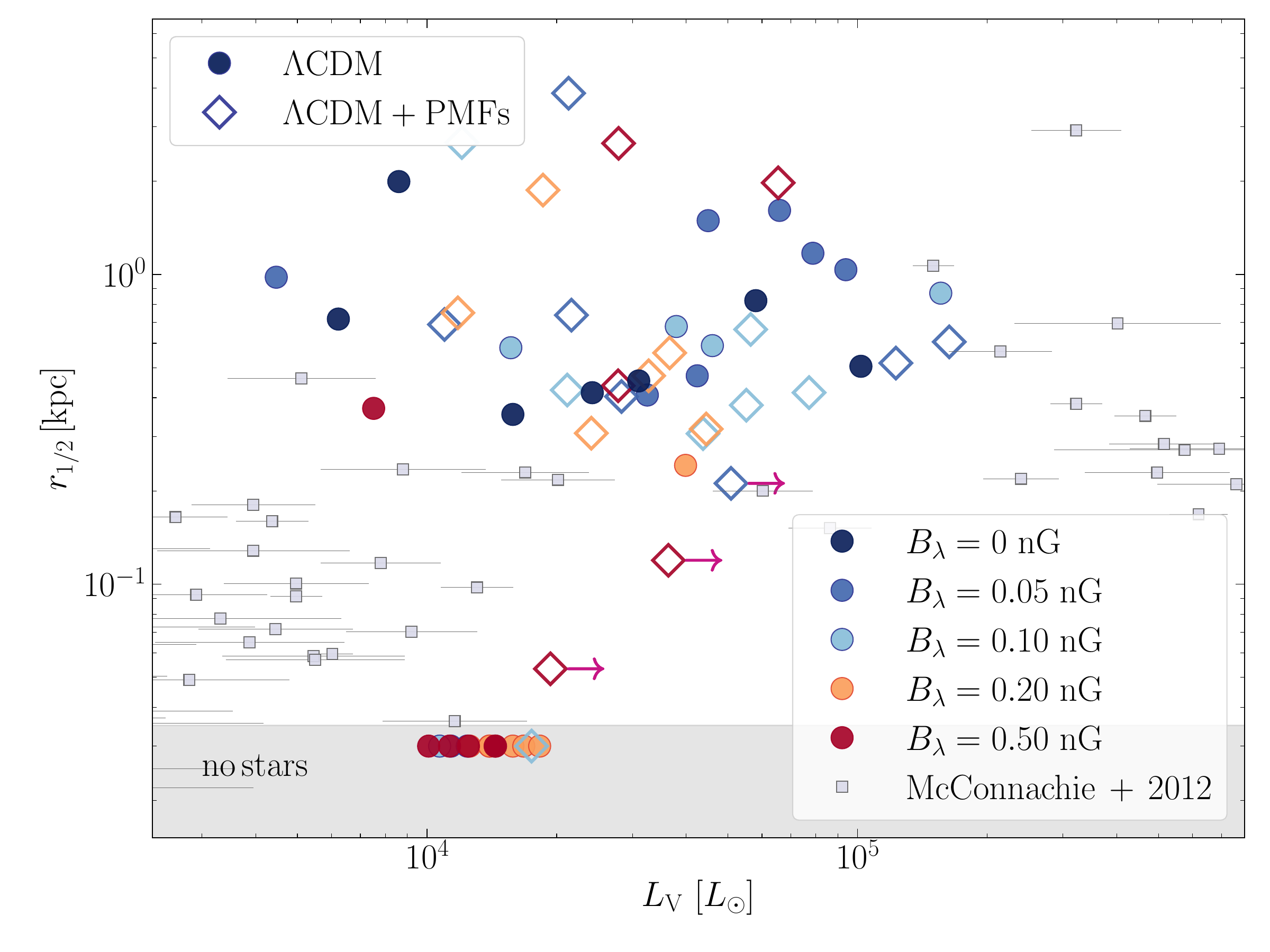}
    \caption{Half-light radius as a function of V-band 
    luminosity
    of each model galaxy at redshift $z=0$ is compared to Milky Way and M31 satellites in black squares, given by \cite{mcconnachie2012}. In order to include six halos which do not form any stars in the plot, their luminosities are shown at $L_{\rm{V}} \sim 10^{3}\,\rm{L_{\odot}}$.} 
    \label{fig:r12}
\end{figure}
%
Figure~\ref{fig:r12} displays the half-light radius $r_{1/2}$ of the model dwarf galaxies as a function of their V-band luminosity at redshift $z=0$.
The Local Group dwarfs compiled from \citet{mcconnachie2012} are represented by gray squares with error bars.
The galaxies simulated in the fiducial model cover a wide range in luminosity and size. 
The half-light sizes in this model span approximately from $300$ to $850\,\mathrm{pc}$.

Galaxies in the first set of simulations, \texttt{$\Lambda$CDM}, are represented by colored circles.
Here, the influence of magnetic fields 
correlates with the half-light radius of the simulated galaxies. The more compact the galaxies, the less effect magnetic fields have on their gas distribution and, consequently, on their star formation. A magnetic field strength of $B_{\lambda}=0.05\,\mathrm{nG}$ in the majority of galaxies in our sample increases the half-light radius by an average value of $25$ percent. However, for $B_{\lambda}>0.05\,\mathrm{nG}$, magnetic fields hinder star formation in small minihalos and by preventing the formation of stars in the outskirts, shrink the size of ultra-faint dwarf galaxies.
It demonstrates that incorporating magnetic fields of strength $B_{\lambda}>0.05\,\mathrm{nG}$ and $B_{\lambda}<0.50\,\mathrm{nG}$ in hydrodynamical simulations leads to a reduction in the size of dwarf galaxies, offering a plausible mechanism for the formation of very compact ultra-faint dwarfs (See e.g., \citet{2023A&A...679A...2R} for a detailed discussion on reproducing compact ultra-faint dwarf galaxies).

Simulated dwarfs in the second set of simulations, \texttt{$\Lambda$CDM+PMFs}, are represented by colored empty diamonds.
In these models, the stellar half-mass is more widely dispersed within the halo of each galaxy. This is a direct consequence of the primordial magnetic fields, which increase the stellar line-of-sight velocity dispersion.
Here, stars form in more abundant subhalos and will be brought to the main halo of the galaxy by mergers. This results in an increased kinematic energy component and higher velocity dispersion \citep{2023A&A...679A...2R}. The large angular momenta of these merging subhalos imply larger sizes for the simulated dwarf galaxies. 
This effect becomes especially pronounced when $B_{\lambda} = 0.20$ and $0.50\,\mathrm{nG}$.

\paragraph{}

To summarize, the $z=0$ properties of halos formed in our fiducial model, provide a good representation of spheroidal and ultra-faint dwarf galaxies in the Local Group, serving as a robust basis for predicting the influence of magnetic fields.
From our findings, we can affirm that primordial magnetic fields with strengths $B_{\lambda}\geq0.05\,\mathrm{nG}$ significantly impact the physical observables of dwarf 
galaxies, specially the small mass ultra-faints with luminosities below $10^5\,L_{\odot}$. 
In this work, the best consistency with the scaling relations of 
Local Group 
dwarfs in our simulated galaxies is obtained when incorporating primordial magnetic fields with moderate strengths of $B_{\lambda}=0.05$ and $0.10\,\mathrm{nG}$ in the MHD configuration.

\paragraph{}


\section{Discussion \& Conclusions}\label{sec:conclusions}


In this study, we investigated the impact of primordial magnetic fields on the formation and evolution of dwarf galaxies.
We conducted new cosmological zoom-in magneto-hydrodynamical simulations from redshift $z=200$ to zero. 
These simulations are focused on eight halos that host dwarfs with V-band luminosities ranging from approximately $10^3$ to  $10^6\,{L_\sun}$.
All simulations are generated using our modified version of the constrained transport MHD \texttt{RAMSES} code.  

We employed two different simulation setups.
In the first setup, a uniform magnetic field is initialized with a given strength and the matter power spectrum follows the classical $\Lambda$CDM paradigm. 
In the second setup, we explored the impact of both varying magnetic field strengths and including the effect of a Gaussian random primordial magnetic field on the matter distribution. 
This approach allowed us to examine the combined effect of primordial magnetic fields on the matter power spectrum, on the one hand, and the influence of 
magneto-hydrodynamics on the evolution of small mass galaxies, on the other hand.

Depending on the strength, each magnetic field model affects a distinct mass range of dark matter halos, within the mass range relevant to host halos of dwarf galaxies. 
In this study, we explored a range of primordial magnetic fields with varying strengths, specifically in the range $B_{\lambda}=0.05-0.50\,\mathrm{nG}$, while keeping the spectral index constant at $n_B=-2.9$. All of these magnetic field strengths fall within the limits allowed by cosmological constraints listed in 
Section~\ref{sec:intro}.
Our main results can be summarized as follows:

\begin{itemize}
    
\item In the context of a $\Lambda$CDM cosmological framework, magnetic fields with an initial strength of $B_{\lambda}\geq0.05\,\mathrm{nG}$, have a non-negligible contribution in shaping the evolution of dwarf galaxies. 
These fields not only delay the collapse of gas clouds, but in the presence of $B_{\lambda}\geq0.20\,\mathrm{nG}$ prevent the formation of dense gas minihalos altogether.
This effect is stronger in low mass ultra-faint dwarf galaxies with halo mass and stellar mass below $M_{200}\sim 2.5\cdot10^9\,M_{\odot}$ and $M_{\star}\sim 3\cdot10^6\,M_{\odot}$, respectively, due to the shallow potential well of their host halos.

\item 
In a cosmological setup evolved from initial conditions of \texttt{$\Lambda$CDM+PMFs}, 
the main halo of a dwarf galaxy, as well as the number of its subhalos, are influenced.
When the halos affected by primordial magnetic fields fall within the Jeans mass range $M_{l}\leq10^8\,\mathrm{M_{\odot}}$, the gas is distributed among a higher number of these low-mass subhalos. On the other hand, if the halo mass range affected contains the Jeans mass of the main halo of a dwarf galaxy ($M_{l}\simeq10^9\,\mathrm{M_{\odot}}$), the gas content in the central halo increases. This strong gravitational potential can lead to the merging of satellite subhalos and ultimately increase the final mass of the host galaxy.

\item
Following the delay caused by magnetic fields  
during the formation of first stars, the gravitational field of low-mass dark matter halos is unlikely to retain gas when it is pushed out by UV-radiation during the epoch of reionisation.  
This results in the depletion of the gas reservoir in minihalos and an earlier quenching of star formation.
In our sample, 
the density and temperature of gas in all six ultra-faint halos remain below the MTT criteria to form stellar particles.
%
However, the influence of primordial magnetic fields on the matter power spectrum accelerates the formation of the first dark matter minihalos, leading to an earlier onset and a higher rate of star formation at redshifts $z>12$.



\item
The observable properties of dwarf galaxies at redshift $z=0$ are indicative of the influence of magnetic fields on their star formation history and kinematics. 
Comparing the two simulation setups, we observe that
dwarf galaxies forming in a \texttt{$\Lambda$CDM+PMFs} Universe with a moderate primordial magnetic field of $B_{\lambda}=0.05$ and $0.10\,\mathrm{nG}$ are more consistent with the scaling relations of
Local Group observations. 
However, stronger magnetic fields lead to dwarf galaxies with larger size and higher velocity dispersion compared to the observed Milky Way satellites.

\item
Increasing the strength of magnetic fields does not result in a proportional $B_{\lambda}^2/8\pi$ increase in magnetic energy. 
This suggests energy loss due to magnetic field saturation, with some energy dissipation even occurring during the collapse phase. The saturation of magnetic energy occurs more rapidly when $B_{\lambda}\geq0.20\,\mathrm{nG}$, within relatively short time intervals of a few hundred $\mathrm{Myrs}$. When $B_{\lambda}\leq0.10\,\mathrm{nG}$, it takes a longer time for the magnetic energy to reach saturation.

\item
In our simulations, the buildup of magnetic energy in time differs from the evolution of magnetic fields in an adiabatic spherical 
collapse with magnetic flux freezing,
also referred to as compressional amplification.
This latter is not driven by the supernovae feedback, suggesting that the amplification of the magnetic energy is not provoked in the course of supernovae explosions.  
Instead, it supports the growth of magnetic fields initialized with a seed as small as $B_{\lambda}=0.05\,\mathrm{nG}$ due to dynamo amplification.

\end{itemize}

In summary, primordial magnetic fields have a substantial impact on the initial stages of dwarf galaxy formation by modifying the matter power spectrum. Moreover, magnetic fields play an important role in the evolution of dwarf galaxies by contributing to the ISM energy budget.
These effects situate dwarf galaxies as an unprecedented window into primordial magnetism. Therefore, the interplay between simulations and observations of spheroidal and ultra-faint dwarf galaxy populations has the potential to further constrain and inform our understanding of the primordial magnetic fields in our Universe. 
On the other hand, the early structure formation induced by primordial magnetic fields is potentially capable of producing rapid and efficient star formation at high redshifts suggested by the JWST data.

\begin{acknowledgements}
This work was performed using the DiRAC Data Intensive service at Leicester, operated by the University of Leicester IT Services, which forms part of the STFC DiRAC HPC Facility (www.dirac.ac.uk). The equipment was funded by BEIS capital funding via STFC capital grants ST/K000373/1 and ST/R002363/1 and STFC DiRAC Operations grant ST/R001014/1. DiRAC is part of the National e-Infrastructure.
J.S.~acknowledges the support from the Swiss National Science Foundation under Grant No.\ 185863.
S.M.A. acknowledges support from the Kavli Institute for Particle Astrophysics and Cosmology (KIPAC) Fellowship.

\end{acknowledgements}

\bibliographystyle{aa}
\bibliography{bibliography}

\end{document}